\documentclass[pre,twocolumn,showpacs,floatfix]{revtex4-1}

\usepackage{amssymb,amsmath}

\usepackage{epsfig}

\usepackage{graphicx}% Include figure files
\usepackage{color}

\newcommand{\emmanuel}[1]{{ #1}}
\newcommand{\vicente}[1]{{ #1}}

\begin{document}
\title{Impurity in a sheared inelastic Maxwell gas}
\author{Vicente Garz\'{o}}
\email{vicenteg@unex.es}
\homepage{URL: http://www.unex.es/eweb/fisteor/vicente/}
\affiliation{Departamento de F\'{\i}sica, Universidad de Extremadura, E-06071 Badajoz, Spain}
\author{Emmanuel Trizac}
\email{trizac@lptms.u-psud.fr}
\homepage{URL: http://www.lptms.u-psud.fr/membres/trizac/}
\affiliation{Laboratoire de Physique
Th\'eorique et Mod\`eles Statistiques (CNRS UMR 8626), B$\hat{a}$timent 100,
Universit\'e Paris-Sud, 91405 Orsay cedex, France}

\begin{abstract}
The Boltzmann equation for inelastic Maxwell models is considered in order to
investigate the dynamics of an impurity (or intruder) immersed
in a granular gas driven by a uniform shear flow.
The analysis is based on an exact solution of the Boltzmann equation for a granular binary mixture.
It applies for conditions arbitrarily far from equilibrium (arbitrary values of the shear rate $a$) and
for arbitrary values of the parameters of the mixture (particle masses $m_i$, mole fractions $x_i$,
and coefficients of restitution $\alpha_{ij}$).
In the tracer limit where the mole fraction of the intruder species vanishes,
a non equilibrium phase transition takes place.
We thereby identity ordered phases where
the intruder bears a finite contribution to the
properties of the mixture, in a region of parameter space that is worked
out in detail.
These findings extend previous results obtained for ordinary Maxwell
gases, and further show that dissipation leads to new ordered phases.
\end{abstract}

\date{\today}
\maketitle

\section{Introduction}
\label{sec1}

Inelastic hard spheres (IHS) provide a useful
theoretical, computational, and experimental framework for studying granular
gases \cite{BP04,BTE05,AT06}.
In the simplest version, the spheres are assumed to be smooth (i.e.
frictionless) and the inelasticity in collisions is specified in terms of
a constant coefficient of normal restitution $\alpha\leq 1$ \cite{BP04}.
At a kinetic theory level, the essential information about the dynamical
properties of the fluid is embedded in the one-particle velocity
distribution function. For sufficiently low-densities, the conventional
Boltzmann equation can be extended to IHS by changing the collision
rules to account for the inelastic character of the interactions
\cite{GS95,BDS97}. However, the complex mathematical structure of the
Boltzmann collision operator prevents one from obtaining exact results, so
that most of the analytical progress for IHS has been achieved by using
approximated methods and/or simple kinetic models \cite{kineticmodels}.
%of the Boltzmann equation.
For instance, the explicit forms of the Navier-Stokes transport coefficients
have been approximately determined by considering the leading terms in a
Sonine polynomial expansion \cite{NS}.

Needless to say, the difficulties of solving the Boltzmann equation increase
considerably when one considers far from equilibrium situations, such as shear
flow problems. Although good estimates for \emph{nonlinear} transport properties of
IHS have been obtained \cite{IHS}, the search of \emph{exact} results has stimulated
the use of model interactions simpler than hard spheres. One possibility is to consider
a mean field version of the hard sphere system, where randomly chosen pairs of particles
inelastically interact with a random impact direction. This assumption yields a
Boltzmann equation where the collision rate is \emph{independent} of the relative
velocity of the two colliding particles \cite{Maxwell}. This interaction model is
usually referred to as the inelastic Maxwell model (IMM). It must be noted that
in the conventional case of ordinary gases colliding elastically, Maxwell models
correspond --in three dimensions-- to particles interacting via a repulsive potential proportional to
the inverse fourth power of distance \cite{CC70}. Nevertheless, in the framework
of the Boltzmann equation, one can introduce Maxwell models at the level of the
cross section, without any reference to a specific interaction potential
\cite{E81,etb06}.
Thanks to the simplifications introduced by IMM in the kernel of the Boltzmann
collision operator, it is possible in some particular situations to find
non-trivial exact solutions to the Boltzmann equation \cite{G03,TK03,SG07,G07,SGV10,GT10}.
Apart from their academic interest, it has also been shown in some cases that the
results derived from IMM \cite{G03} agree well with those obtained analytically for
IHS from Grad's method \cite{MG02,SGD04} and by means of Monte Carlo simulations
\cite{MG02}. We will substantiate this point in the concluding section.
In addition, even recent experiments \cite{exp05} for
magnetic grains with dipolar interactions are qualitatively well described
by IMM. All of these results clearly show the utility of IMM as a toy model to
unveil in a clean way the influence of the inelasticity of collisions in granular flows,
especially in far from equilibrium situations where simple intuition is not enough.

The aim of this paper is to investigate the dynamics of an intruder or
impurity immersed in an inelastic Maxwell gas subject to the simple or
uniform shear flow (USF). This state is perhaps one of the most widely
studied states in granular gases \cite{Gol03}. From a macroscopic point
of view, the USF is characterized by constant partial densities $n_r$, a
uniform granular temperature $T$, and a linear velocity field $u_x=ay$, $a$
being the constant shear rate. Because the only hydrodynamic gradient is
that of flow velocity, the mass and heat fluxes vanish by symmetry and the
pressure or stress tensor $P_{ij}$ is the only relevant flux in the problem.
The knowledge of the elements of $P_{ij}$ gives access to the rheological
properties of the mixture. In the context of IMM, the above transport
properties have been recently obtained \cite{GT10} in terms of the shear
rate and the parameters of the mixture. Our goal now is to consider the
tracer limit. Since this case corresponds to a situation in which
the \emmanuel{mole fraction $x_1=n_1/(n_1+n_2)$ of the tracer species} is negligible,
one expects that the properties of the excess species (granular gas) are
not affected by the presence of the impurity particle. In particular, the
relative contribution of the impurity to the total energy of the system,
$E_1/E$, is expected to be proportional to $x_1$ when $x_1 \ll 1$
\emmanuel{(here, $E$ denotes the total kinetic energy of the system and
$E_1$ is the kinetic energy of the impurity)}.
Consequently, the contribution of the impurity to the total energy is
likely to be negligible. However, as in the elastic case \cite{MSG96},
we present in this paper a violation of the above expectation, that is
ascribable to a non-equilibrium phase transition.
Specifically, we found that
the impurity particle has a \emph{finite} contribution to the total energy
of the system (impurity plus granular gas) either for light impurities under
conditions that depend critically on the asymmetry of collisional dissipation
(impurity-gas against gas-gas collisions), or conversely, for heavy impurities
provided the shear rates are not too large. This phenomenon
is akin to a phase transition, where the order parameter is
given by the ratio of the impurity kinetic energy over the total
kinetic energy of the system. \emmanuel{We borrow here the terminology
of equilibrium phase transitions; it should nevertheless be stressed that
the kind of ordering under discussion is not spatial, but refers to
a specific, kinetic, order parameter ($E_1/E$); as such, the ``ordering'' scenario
specifically belongs in non-equilibrium.}
Moreover, we found that ordering always sets in
at large enough shear rates (i.e., for shear rates larger than a certain critical value),
provided the impurity is sufficiently light compared to the grains of the host gas.
\emmanuel{While this feature is common to elastic \cite{MSG96} and inelastic systems,
the possibility of what we refer below as heavy tracer ordering, is
specific to inelastic gases.} A preliminary account of part of this work has appeared in Ref.\ \cite{GT11}.

The plan of the paper is as follows. The Boltzmann equation for IMM is
introduced in Sec.\ \ref{sec2} and the collisional moments needed to
get the pressure tensor are explicitly evaluated. In Sec.\ \ref{sec3},
the USF problem is presented and the rheological properties are obtained
in terms of the shear rate, the masses, the concentration, and the coefficients of restitution.
The main results are derived in Sec.\ \ref{sec4}. Specifically, the tracer
limit is considered in detail, which shows the existence of the above
non-equilibrium transition. Finally, the paper is closed in
Sec.\ \ref{sec5} with a brief discussion.

\section{The inelastic Maxwell model}
\label{sec2}

Let us consider a binary mixture of inelastic Maxwell gases at low density. In the
absence of external forces, the set of nonlinear  Boltzmann equations for the mixture
reads
\begin{equation}
\label{2.1}
\left(\frac{\partial}{\partial t}+{\bf v}\cdot \nabla \right)f_{r}
({\bf r},{\bf v};t)
=\sum_{s}J_{rs}\left[{\bf v}|f_{r}(t),f_{s}(t)\right] \;,
\end{equation}
where $f_r({\bf r},{\bf v};t)$ is the one-particle distribution function of species $r$ ($r=1,2$) and
the Boltzmann collision operator $J_{rs}\left[{\bf v}|f_{r},f_{s}\right]$ describing the scattering of
pairs of particles is
\begin{widetext}
\begin{equation}
J_{rs}\left[{\bf v}_{1}|f_{r},f_{s}\right]  =\frac{\omega_{rs}}{n_s\Omega_d}
\int d{\bf v}_{2}\int d\widehat{\boldsymbol {\sigma }}\left[ \alpha_{rs}^{-1}f_{r}({\bf r},{\bf v}_{1}',t)f_{s}(
{\bf r},{\bf v}_{2}',t)-f_{r}({\bf r},{\bf v}_{1},t)f_{s}({\bf r},{\bf v}_{2},t)\right]
\;.
\label{2.2}
\end{equation}
\end{widetext}
Here,
\begin{equation}
\label{2.4.1} n_r=\int d{\bf v} f_r({\bf v})
\end{equation}
is the number density of species $r$, $\omega_{rs}$ is an effective collision frequency
(to be chosen later) for collisions  of type $r$-$s$,  $\Omega_d=2\pi^{d/2}/\Gamma(d/2)$
is the total solid angle in $d$ dimensions, and $\alpha_{rs}\leq 1$ refers to the
constant coefficient of restitution  for collisions between particles of species $r$
with $s$.   In addition, the primes on the velocities denote the initial values $\{{\bf
v}_{1}^{\prime}, {\bf v}_{2}^{\prime}\}$ that lead to $\{{\bf v}_{1},{\bf v}_{2}\}$
following a binary collision:
\begin{equation}
\label{2.3}
{\bf v}_{1}^{\prime }={\bf v}_{1}-\mu_{sr}\left( 1+\alpha_{rs}
^{-1}\right)(\widehat{\boldsymbol {\sigma}}\cdot {\bf g}_{12})\widehat{\boldsymbol
{\sigma}},
\end{equation}
\begin{equation}
\label{2.3.1}
{\bf v}_{2}^{\prime}={\bf v}_{2}+\mu_{rs}\left(
1+\alpha_{rs}^{-1}\right) (\widehat{\boldsymbol {\sigma}}\cdot {\bf
g}_{12})\widehat{\boldsymbol{\sigma}}\;,
\end{equation}
where ${\bf g}_{12}={\bf v}_1-{\bf v}_2$ is the relative velocity of the colliding pair,
$\widehat{\boldsymbol {\sigma}}$ is a unit vector directed along the centers of the two colliding
spheres, and $\mu_{rs}=m_r/(m_r+m_s)$. From the densities $n_r$, we define
the mole fractions $x_r = n_r/(n_1+n_2)$.

The effective collision frequencies $\omega_{rs}$ are independent of the relative velocities
of the colliding particles but depend \emmanuel{in general}
on space and time through its dependence on density and temperature. In previous works on multicomponent granular systems \cite{G03,GT10,GA05}, $\omega_{rs}$ was chosen to guarantee that the cooling rate for IMM be the same as that of the IHS. With this choice (``improved Maxwell model''), the collision rates $\omega_{rs}$
are functions of the temperature ratio $\gamma\equiv T_1/T_2$, that is itself a
function of the (reduced) shear rate in the USF problem. A consequence of this
choice is that one has to \emph{numerically} solve a set of nonlinear equations
in order to get the shear rate dependence of the temperature ratio \cite{G03,GT10}.
Since we wish to obtain in this paper analytical results for arbitrary spatial
dimensions in a quite complex problem that involves a delicate tracer limit,
we will consider here a simple version of IMM where $\omega_{rs}$ is independent of the partial
temperatures of each species (``plain vanilla Maxwell model''). Thus, one defines $\omega_{rs}$ as
\begin{equation}
\omega_{rs}=x_s\nu_0, \quad \nu_0=A n,
\end{equation}
where the value of the constant $A$ is irrelevant for our purposes.
Here, $n=\sum_r n_r$ is the total number density of the mixture.
The form of $\omega_{rs}$ is closer to the
original model of Maxwell molecules for elastic mixtures.
This plain vanilla model has been previously used by several authors \cite{MP02a,MP02b,NK02,CMP07}
in some problems pertaining to granular mixtures, and we will argue in section \ref{sec5}
that it is capable of capturing the essential physical effects at work here.

At a hydrodynamic level,  the relevant quantities in a binary mixture
are, apart from $n_r$, the flow velocity  ${\bf u}$, and the ``granular'' temperature $T$. They are
defined in terms of moments of the distribution $f_r$ as
\begin{equation}
\label{2.4}
\rho{\bf u}=\sum_r\rho_r{\bf u}_r=\sum_r\int d{\bf v}m_r{\bf v}f_r({\bf
v}),
\end{equation}
\begin{equation}
\label{2.5}
nT=\sum_rn_rT_r=\sum_r\int d{\bf v}\frac{m_r}{d}V^2f_r({\bf v}),
\end{equation}
where $\rho_r=m_rn_r$, $\rho=\sum_r \rho_r$ is
the total mass density, and ${\bf V}={\bf v}-{\bf u}$ is the peculiar velocity.
Equations (\ref{2.4}) and (\ref{2.5}) also define the flow velocity ${\bf u}_r$ and the
partial temperature $T_r$ of species $r$, the latter measuring the mean kinetic energy
of species $r$. Computer simulations \cite{computer}, experiments
\cite{exp} and kinetic theory calculations \cite{JM87,GD99} indicate that the global granular temperature
$T$ is in general different from the partial temperatures $T_r$. The mass,
momentum and energy fluxes are characterized by the mass flux
\begin{equation}
{\bf j}_{r}=m_{r}\int d{\bf v}\,{\bf V}\,f_{r}({\bf v}),
\label{2.6}
\end{equation}
the pressure tensor
\begin{equation}
{\sf P}=\sum_{r}\,\int d{\bf v}\,m_{r}{\bf V}{\bf V}\,f_{r}({\bf  v}),
\label{2.7}
\end{equation}
and the heat flux
\begin{equation}
{\bf q}=\sum_{r}\,\int d{\bf v}\,\frac{1}{2}m_{r}V^{2}{\bf V}
\,f_{r}({\bf v}),
\label{2.8}
\end{equation}
respectively. Finally, the rate of energy dissipated due to collisions among all species defines the cooling rate $\zeta$ as
\begin{equation}
\sum_{r,s}m_r\int d{\bf v}V^{2}J_{rs}[{\bf v}
|f_{r},f_{s}]=-d nT\zeta \;.  \label{2.9}
\end{equation}

The key advantage of the Boltzmann equation for Maxwell models (both elastic and
inelastic) is that the (collisional) moments of $J_{rs}[f_r,f_s]$ can be exactly
evaluated in terms of the moments of $f_r$ and $f_s$ without the explicit knowledge of
both velocity distribution functions \cite{TM80,GS03}. This property has been recently exploited
\cite{GS07} to obtain the detailed expressions for all the second-, third- and
fourth-degree collisional moments for a monodisperse gas. In the case of a binary
mixture, only the first- and second-degree collisional moments have been also
explicitly obtained. In particular \cite{G03},
%\begin{widetext}
\begin{eqnarray}
\label{2.10}
\int d{\bf v} m_r {\bf V} {\bf V}J_{rs}[f_r,f_s]&=&
-\frac{\omega_{rs}}{\rho_sd}\mu_{sr}(1+\alpha _{rs})\left\{2\rho_s{\sf P}_r\right.\nonumber\\
&-&\left(
{\bf j}_r{\bf j}_s+{\bf j}_s{\bf j}_r\right)-\frac{2}{d+2}\mu_{sr}(1+\alpha _{rs})\nonumber\\
&\times& \left[\rho_s{\sf P}_r+\rho_r{\sf P}_s
-\left({\bf j}_r{\bf j}_s+{\bf j}_s{\bf j}_r\right)\right.\nonumber\\
&+ & \left.\left.
 \left[\frac{d}{2}\left(\rho_rp_s+\rho_sp_r\right)
-{\bf j}_r\cdot {\bf j}_s\right]\openone\right]\right\},
\nonumber\\
\end{eqnarray}
%\end{widetext}
where
\begin{equation}
\label{2.11}
{\sf P}_r=\int d{\bf v}\,m_{r}{\bf V}{\bf V}\,f_{r},
\end{equation}
$p_r=n_rT_r=\text{tr}{\sf P}_{r}/d$ is the partial pressure of species $r$, and
$\openone$ is the $d\times d$ unit tensor. It must be remarked that, in general beyond
the linear hydrodynamic regime (Navier-Stokes order), the above property of the
Boltzmann collision operator is not sufficient to exactly solve the hierarchy of moment
equations, due to the free-streaming term of the Boltzmann equation. Nevertheless, there
exist some particular situations (such as the simple shear flow problem) for which the
above hierarchy can be recursively solved. The cooling rate $\zeta$ defined by Eq.\ (\ref{2.9}) can be easily obtained from Eq.\ (\ref{2.10}) as
\begin{widetext}
\begin{equation}
\label{2.12}
\zeta=\frac{2}{d}\sum_{r,s}x_r\omega_{rs}\mu_{sr}(1+\alpha _{rs})\left[\gamma_r-\frac{1+\alpha_{rs}}{2}
(\gamma_r\mu_{sr}+\gamma_s\mu_{rs})
+\gamma_r\frac{\mu_{sr}(1+\alpha _{rs})-1}{d\rho_sp_r}
{\bf j}_r\cdot {\bf j}_s\right],
\end{equation}
\end{widetext}
where $\gamma_r\equiv T_r/T$.

\section{Uniform shear flow}
\label{sec3}

We consider a binary mixture of inelastic Maxwell gases under USF.
As mentioned in the Introduction, the USF state is
macroscopically defined by constant densities $n_r$, a spatially uniform temperature
$T(t)$ and a linear velocity profile ${\bf u}(y)={\bf u}_1(y)={\bf u}_2(y)=ay \,
\widehat{\mathbf{x}}$, where $a$ is the {\em constant} shear rate. Since $n_r$ and $T$
are uniform, then ${\bf j}_r={\bf q}={\bf 0}$, and the transport of momentum (measured
by the pressure tensor) is the relevant phenomenon. At a microscopic level, the USF is
characterized by a velocity distribution function that becomes {\em uniform} in the
local Lagrangian frame, i.e., $f_r({\bf r},{\bf v};t)=f_r({\bf V},t)$. In this frame,
the Boltzmann equation (\ref{2.1}) for the distribution $f_1({\bf V},t)$ reads \cite{GS03}
\begin{equation}
\label{3.1} \frac{\partial}{\partial t}f_1-aV_y\frac{\partial}{\partial
V_x}f_1=J_{11}[f_1,f_1]+J_{12}[f_1,f_2] ,
\end{equation}
while a similar equation holds for $f_2$. The properties of uniform temperature and constant densities and shear rate are
enforced in computer simulations by applying the Lees-Edwards boundary conditions
\cite{LE72}, regardless of the particular interaction model considered. In the case of
boundary conditions representing realistic plates in relative motion, the corresponding non-equilibrium state
is the so-called Couette flow, where densities, temperature and shear rate are
no longer uniform \cite{Tij01,VSG10}.

As alluded to above, the rheological properties of the mixture are obtained from the
pressure tensor ${\sf P}={\sf P}_1+{\sf P}_2$, where the partial pressure tensors ${\sf
P}_r$ ($r=1,2$) are defined by Eq.\ (\ref{2.11}). The elements of these tensors can be
obtained by multiplying the Boltzmann equation (\ref{3.1}) by $m_r{\bf V}{\bf V}$ and
integrating over ${\bf V}$. The result can be written as
\begin{widetext}
\begin{equation}
\label{3.5}
\frac{\partial}{\partial t}P_{1,ij}+a_{ik}P_{1,kj}+a_{jk}P_{1,ki}+B_{11}P_{1,ij}+B_{12}P_{2,ij}=
\left(A_{11}p_{1}+A_{12}p_2\right)\delta_{ij},
\end{equation}
\end{widetext}
where use has been made of Eq.\ (\ref{2.10}) (with ${\bf j}_r={\bf 0}$). In Eq.\
(\ref{3.5}), $a_{ij}=a\delta_{ix}\delta_{jy}$ and we have introduced the coefficients
\begin{equation}
\label{3.6}
A_{11}=\frac{\omega_{11}}{2(d+2)}(1+\alpha_{11})^2+\frac{\omega_{12}}{d+2}\mu_{21}^2(1+\alpha_{12})^2,
\end{equation}
\begin{equation}
\label{3.7}
A_{12}=\frac{\omega_{12}}{d+2}\frac{\rho_1}{\rho_2}\mu_{21}^2(1+\alpha_{12})^2,
\end{equation}
\begin{eqnarray}
\label{3.8}
B_{11}&=&\frac{\omega_{11}}{d(d+2)}(1+\alpha_{11})(d+1-\alpha_{11})
+\frac{2\omega_{12}}{d(d+2)}\nonumber\\
& & \times \mu_{21}(1+\alpha_{12})
\left[d+2-\mu_{21}(1+\alpha_{12})\right],
\end{eqnarray}
\begin{equation}
\label{3.9}
B_{12}=-\frac{2}{d}A_{12}.
\end{equation}
Adequate change of indices ($1\leftrightarrow 2$) provide the equations pertaining to
${\sf P}_2$. The evolution equation for the temperature  can be obtained
from Eq.\ (\ref{3.5}) and reads
\begin{equation}
\label{3.10}
\nu_0^{-1}\frac{\partial}{\partial t}\ln T=-\zeta^*-\frac{2a^*}{d} P_{xy}^*.
\end{equation}
Here, the following reduced quantities have been introduced:
$\zeta^*=\zeta/\nu_0$, $a^*=a/\nu_0$, $P_{xy}^*=P_{xy}/p$, $p=nT$ being the
hydrostatic pressure. The expression for $\zeta^*$ can be derived from Eq.\
(\ref{2.12}) when one takes ${\bf j}_r={\bf 0}$. The result is
\begin{equation}
\label{3.10.1}
\zeta^*=\frac{2}{d}\sum_{r,s}\; x_rx_s
\mu_{sr}(1+\alpha_{sr})
\left[\gamma_r-\frac{1+\alpha_{rs}}{2}(\gamma_r\mu_{sr}+\gamma_s\mu_{rs})
\right].
\end{equation}

As can be seen in Eq.\ (\ref{3.10}), the temperature changes in time due the competition of two
opposite mechanisms: viscous heating (shearing work) and energy dissipation
in collisions. The {\em reduced} shear rate $a^*$ is
the non-equilibrium relevant parameter of the USF problem since it measures the distance
of the system from the homogeneous cooling state \cite{BP04}. In general, since $a^*$ does not depend on time, there is no steady state unless $a^*$ takes the specific value given by the steady-state condition
\begin{equation}
\label{3.11}
a_s^*P_{s,xy}^*=-\frac{d}{2}\zeta^*,
\end{equation}
denoting $a_s^*$ and $P_{s,xy}^*$ the steady-state values of the (reduced) shear rate
and the pressure tensor. Beyond this particular case, the (reduced) shear rate and the
coefficients of restitution can be considered as independent and so, one
can analyze the combined effect of both control parameters on the rheological properties
of the mixture. This is one of the main advantages of the interaction model used in this
paper, in contrast to previous works \cite{G03}\emmanuel{: our goal
is to disentangle the effects of dissipation from those of
forcing through shear. For comparison to a real
system or with simulation data, that are generally studied in the steady
state where collisional cooling and viscous heating equilibrate,
one should however specifically work with an $\alpha$-dependent (reduced) shear rate, given by the solution of Eq. (\ref{3.11}). We shall come back to this point in Sec.\ \ref{sec5}.}

\begin{figure}
\includegraphics[width=0.9\columnwidth]{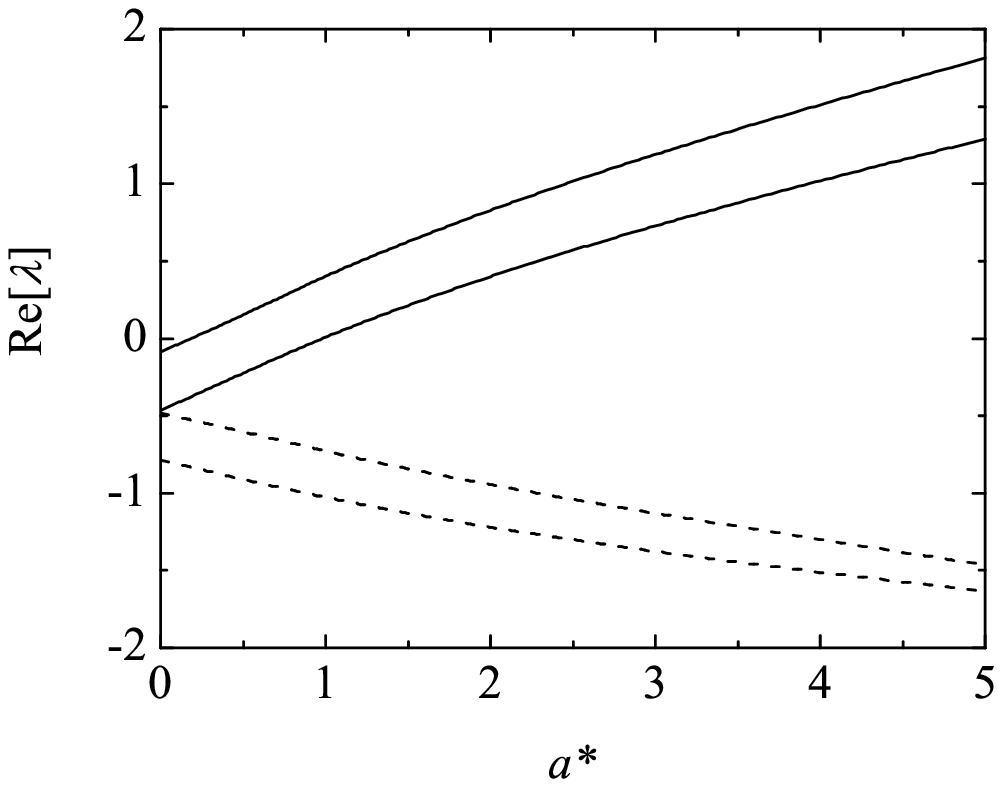}
\caption{Shear rate dependence of the real part of the roots of Eq.\  (\ref{3.15}) for
hard disks ($d=2$) in the case $x_1=0.2$, $m_1/m_2=0.5$, and
$\alpha_{11}= \alpha_{12}=\alpha_{22}=0.8$. The
solid lines refer to the real roots, while the dashed lines refer to the complex roots.
}\label{Fig:1}
\end{figure}

We are interested in obtaining the explicit forms of the scaled pressure tensors
$P_{r,ij}^*=P_{r,ij}/p$ in the long-time limit. The relevant elements of these tensors are $P_{r,xx}^*=p_r^*-(d-1)P_{r,yy}^*$, $P_{r,yy}^*$, and $P_{r,xy}^*$ with $r=1,2$ \cite{G03,GT10}. Here, $p_r^*=p_r/p=x_r\gamma_r$. As in the monocomponent granular case \cite{SG07}, one can check that, after a
certain kinetic regime lasting a few collision times, the scaled pressure tensors
$P_{r,ij}^*=P_{r,ij}/p$ reach well-defined stationary values
(non-Newtonian hydrodynamic regime), which are non-linear
functions of the (reduced) shear rate $a^*$ and the coefficients of
restitution. In terms of these scaled variables and by using matrix notation, Eq.\
(\ref{3.5}) can be rewritten as
\begin{equation}
\label{3}
{\cal L}{\cal P}=0,
\end{equation}
where ${\cal P}$ is the column matrix
\begin{equation}
\label{3.12}
{\cal P}=\left(
\begin{array}{c}
P_{1,xx}^*\\
P_{1,yy}^*\\
P_{1,xy}^*\\
P_{2,xx}^*\\
P_{2,yy}^*\\
P_{2,xy}^*
\end{array}
\right)
\end{equation}
and ${\cal L}$ is the square matrix
\begin{widetext}
\begin{equation}
\label{3.13}
{\cal L}=\left(
\begin{array}{cccccc}
B_{11}^*+\lambda-\frac{1}{d}A_{11}^*&-\frac{d-1}{d}A_{11}^*&2a^*&B_{12}^*-\frac{1}{d}A_{12}^*&
-\frac{d-1}{d}A_{12}^*&0\\
-\frac{1}{d}A_{11}^*&B_{11}^*+\lambda-\frac{d-1}{d}A_{11}^*&0&-\frac{1}{d}A_{12}^*&B_{12}^*
-\frac{d-1}{d}A_{12}^*&0\\
0&a^*&B_{11}^*+\lambda&0&0&B_{12}^*\\
B_{21}^*-\frac{1}{d}A_{21}^*&-\frac{d-1}{d}A_{21}^*&0&B_{22}^*+\lambda-\frac{1}{d}A_{22}^*&-
\frac{d-1}{d}A_{22}^*&2a^*\\
-\frac{1}{d}A_{21}^*&B_{21}^*-\frac{d-1}{d}A_{21}^*&0&-\frac{1}{d}A_{22}^*&B_{22}^*+\lambda-
\frac{d-1}{d}A_{22}^*&0\\
0&0&B_{21}^*&0&a^*&B_{22}^*+\lambda
\end{array}
\right).
\end{equation}
\end{widetext}
Here, $A_{rs}^*=A_{rs}/\nu_0$ and $B_{rs}^*=B_{rs}/\nu_0$. In addition,
it has been taken into account that for long times the temperature $T(t)$
behaves as
\begin{equation}
\label{3.14}
T(t)=T(0)e^{\lambda \nu_0 t},
\end{equation}
where $\lambda$ is also a nonlinear function of $a^*$, $\alpha_{rs}$ and the parameters
of the mixture. The (reduced) total pressure tensor $P_{ij}^*=P_{ij}/p$ of the mixture is defined as
\begin{equation}
\label{3.14.1}
P_{ij}^*= P_{1,ij}^*+P_{2,ij}^*.
\end{equation}
Equation (\ref{3}) has a nontrivial solution if
\begin{equation}
\label{3.15}
\det {\cal L}=0.
\end{equation}
Equation (\ref{3.15}) is a sixth-degree polynomial equation with coefficients depending
on $a$ and ${\boldsymbol {\xi}}\equiv \{x_1, \mu, \alpha_{11},
\alpha_{22},\alpha_{12}\}$. Here, $\mu\equiv m_1/m_2$ is the mass ratio. In general, this
equation must be solved numerically.  Figure \ref{Fig:1} shows the real
part $\text{Re}[\lambda]$ of the roots of Eq.\ (\ref{3.15}) versus $a^*$ for hard disks ($d=2$)
in the case $x_1=0.2$, $m_1/m_2=0.5$, and $\alpha_{rs}=0.8$. Obviously, exactly the
same curves are obtained in the case $m_1/m_2=2$ and $x_1=0.8$. At a given value of the
shear rate, the difference between the two largest values of $\nu_0 \lambda$ gives the inverse of the
relaxation time of the transient regime. It can be proved that this difference does not vanish
if $x_1\neq 0$.

According to Eq.\ (\ref{3.14}), the largest root of the sixth-degree equation (\ref{3.15}) governs the time evolution of the global temperature $T(t)$ in the long-time limit. Thus, the upper curve in Fig.\ \ref{Fig:1} gives the value of $\lambda$ of Eq.\ \eqref{3.14} for the case $x_1=0.2$, $m_1/m_2=0.5$, and $\alpha_{rs}=0.8$.

The stationary forms of ${\sf P}_1$ and ${\sf P}_2$ are obtained by solving the
homogeneous equation (\ref{3}) since $\det {\cal L}=0$.
This equation has a nontrivial solution that can be written as
\begin{equation}
\label{7}
{\cal P}'={\cal L}^{'-1}\cdot {\cal Q},
\end{equation}
where
\begin{equation}
\label{8}
{\cal P}'=\left(
\begin{array}{c}
p_1^*\\
P_{1,yy}^*\\
P_{2,yy}^*\\
P_{1,xy}^*\\
P_{2,xy}^*
\end{array}
\right),
\end{equation}
\begin{widetext}
\begin{equation}
\label{9} {\cal L}'=\left(
\begin{array}{ccccc}
B_{11}^*+\lambda-A_{11}^*+A_{12}^*-B_{12}^*&0&0&\frac{2}{d}a^*&0\\
A_{12}^*-A_{11}^*&\lambda+B_{11}^*&B_{12}^*&0&0\\
A_{22}^*-A_{21}^*&B_{21}^*&\lambda+B_{22}^*&0&0\\
0&a^*&0&\lambda+B_{11}^*&B_{12}^*\\
0&0&a^*&B_{21}^*&B_{22}^*+\lambda
\end{array}
\right),
\end{equation}
\end{widetext}
and
\begin{equation}
\label{9.1}
{\cal Q}=\left(
\begin{array}{c}
A_{12}^*-B_{12}^*\\
A_{12}^*\\
A_{22}^*\\
0\\
0
\end{array}
\right).
\end{equation}
The expressions of $p_1^*$ and $P_{r,ij}^*$ can be obtained from Eq.\ (\ref{7}).
In particular, the explicit form of $p_1^*=x_1\gamma_1$
can be found in Appendix \ref{appA}. This quantity gives the ratio between the
energy of the species $1$ and the total energy of the mixture.

It is important to recall that, although the scaled pressure tensors $P_{r,ij}^*$
achieve stationary values, the binary mixture is not in general in a steady state since
the granular temperature changes in time. In fact, since $P_{xy}^*<0$, according to Eq.\ (\ref{3.10})
 the temperature $T(t)$ grows exponentially if $-2a^*P_{xy}^*>d\zeta^*$, namely,
when the imposed shear rate is large enough to make the viscous heating effect
dominate over the collisional cooling. The opposite occurs when
$d\zeta^*>-2a^*P_{xy}^*$ and so, the temperature decreases in time.

\begin{figure}
\includegraphics[width=0.9\columnwidth]{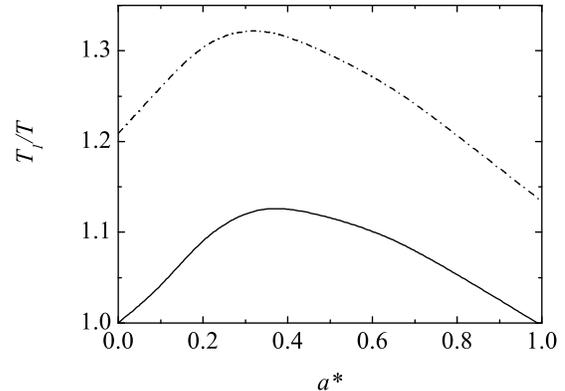}
\caption{Shear rate dependence of the temperature ratio $T_1/T$ in three dimensions
($d=3$), for an equimolar mixture ($x_1=0.5$) with $m_1/m_2=8$. Two values of the
(common) coefficient of restitution $\alpha_{rs}\equiv\alpha$ have been considered:
$\alpha=1$ (solid line) and $\alpha=0.8$ (dashed-dotted line). }\label{Fig:2}
\end{figure}

\begin{figure}
\includegraphics[width=0.9\columnwidth]{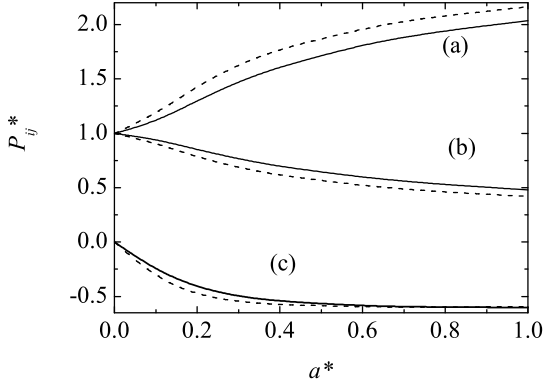}
\caption{Shear rate dependence of the (reduced) elements of the pressure tensor
$P_{xx}^*$ (a), $P_{yy}^*$ (b) and $P_{xy}^*$ (c). Two values of the (common)
coefficient of restitution $\alpha_{rs}\equiv\alpha$ have been considered: $\alpha=1$
(solid lines) and $\alpha=0.8$ (dashed lines). }\label{fig5}
\end{figure}

To illustrate the non-Newtonian behavior \cite{rqueref} of the temperature ratio and the pressure
tensor, Figs.\ \ref{Fig:2} and \ref{fig5} show $\gamma_1$ and $P_{rs}^*$,
respectively, as functions of the (reduced) shear rate $a^*$ for an equimolar mixture
($x_1=0.5$) of inelastic hard spheres ($d=3$) with $m_1/m_2=8$. Two different values of
the (common) coefficient of restitution $\alpha_{rs}\equiv \alpha$ have been
considered. The temperature ratio $T_1/T$ measures the lack of equipartition of kinetic
energy. As expected for driven granular mixtures \cite{MP02b,computer}, the lighter
particles have a smaller temperature for moderate shear rates, while the opposite
happens at high shear rates. Figure \ref{fig5} shows the dependence of the relevant
elements of the total pressure tensor $P_{rs}^*$ on $a^*$. A signal of the
non-Newtonian behavior is the existence of normal stress differences in the shear flow
plane. It is also apparent that the influence of collisional dissipation on the
rheological properties (measured through the elements $P_{rs}^*$) is not quite
significant, especially at high shear rates. It must be remarked that the trends
observed for this plain vanilla IMM turn to be very similar to those previously obtained
from the improved model of IMM \cite{GT10} .

%%%%%%%%%%%%%%%%%%%%%%%%%%%%%%%%%%%%%%%%%%%%%%%%%%%%%%%%%%%%%%%%%%%%%%%%%%%%
\section{Tracer limit ($x_1\to 0$)}
\label{sec4}

The results derived in the preceding section have shown that the time dependent
solution for the second-degree velocity moments is given in terms of the roots of the
sixth-degree polynomial equation (\ref{3.15}). For long times, the dominant behavior is
described by the two real roots, $\lambda_1$ and $\lambda_2$. In particular, the energy
ratio $E_1/E=x_1\gamma_1$ can be written as
\begin{equation}
\label{9.2}
\frac{E_1}{E}=\frac{A p_1^*(\lambda_2,a,\boldsymbol{\xi})+B
p_1^*(\lambda_1,a,\boldsymbol{\xi})e^{-(\lambda_2
-\lambda_1)\nu_0 t}}{A+B e^{-(\lambda_2-\lambda_1)\nu_0 t}},
\end{equation}
where $A$ and $B$ are constants depending on the initial conditions and the explicit
expression of $p_1^*$ is given by Eq.\ (\ref{a5}). After a relaxation
time of the order of $|(\lambda_2-\lambda_1)\nu_0|^{-1}$, the energy ratio $E_1/E$
reaches a steady state value $p_1^*(\lambda_{\text{max}},a,\boldsymbol{\xi})$ where
$\lambda_{\text{max}}=\text{max}(\lambda_2,\lambda_1)$. As long as $x_1\neq 0$, one has
$\lambda_1\neq \lambda_2$ for any value of the shear rate and $\boldsymbol{\xi}$.

Let us assume now that the mole fraction of one of the species (say for instance, species 1) becomes negligible. In the tracer limit ($x_1\to 0$), the sixth-degree
equation (\ref{3.15}) for $\lambda$ factorizes into two cubic equations (see appendix
\ref{appAbis}, where the quantities used below are defined), with the
following largest roots:
\begin{eqnarray}
\label{16}
\lambda_2^{(0)}&=&2F(a^*/A_{22}^{(0)})A_{22}^{(0)}-\left(B_{22}^{(0)}-A_{22}^{(0)}\right)\nonumber\\
&=&
\frac{(1+\alpha_{22})^2}{d+2}F(\widetilde{a})-\frac{1-\alpha_{22}^2}{2d},
\end{eqnarray}
where
\begin{equation}
\label{17.1}
F(x)\equiv \frac{2}{3}\sinh^2\left[\frac{1}{6}\cosh^{-1}\left(1+\frac{27}{d}x^2\right)\right]
\end{equation}
and
\begin{equation}
\label{18}
\widetilde{a}=\frac{a^*}{A_{22}^{(0)}}=\frac{2(d+2)}{(1+\alpha_{22})^2}
a^*.
\end{equation}
The above root $\lambda_2^{(0)}$ rules the dynamics of the host fluid (excess component)
while the evolution of the impurity is governed by the root
\begin{eqnarray}
\label{17}
\lambda_1^{(0)}&=&2F(a^*/A_{11}^{(0)})A_{11}^{(0)}-\left(B_{11}^{(0)}-A_{11}^{(0)}\right)\nonumber\\
&=&
\frac{2\mu_{21}^2}{d+2}(1+\alpha_{12})^2F\left(\frac{\widetilde{a}}{2\mu_{21}^2}
\frac{(1+\alpha_{22})^2}{(1+\alpha_{12})^2}\right)\nonumber\\
& &
-\frac{2}{d}\mu_{21}(1+\alpha_{12})\left[1-\frac{\mu_{21}}{2}(1+\alpha_{12})\right].
\end{eqnarray}

As seen above, the largest of all roots, $\lambda_{\text{max}}$, is the relevant
one to obtain the asymptotic energy ratio $E_1/E$. We now show that the
behavior of the system is qualitatively very different depending on
$\lambda_{\text{max}} = \lambda_1^{(0)}$ or $\lambda_{\text{max}} = \lambda_2^{(0)}$.
For $x_1 \neq 0$, the expression of $p_1^*$ (that corresponds to the long-time value reached
by $E_1/E$)
is given in Appendix \ref{appA}  by Eq.\ (\ref{a5}).
If $x_1 \to 0$, the energy ratio $p_1^*$ becomes
\begin{equation}
\label{20}
p_1^*(\lambda,a^*,{\boldsymbol {\xi}})\approx x_1 \frac{D(\lambda,a^*)}{\Delta_0(\lambda,a^*)+\Delta_1(\lambda,a^*)x_1},
\end{equation}
where the dependence on $\mu$, $\alpha_{11}$, $\alpha_{22}$, and $\alpha_{12}$ is implicitly
assumed on the right-hand side. The general expressions of $D$, $\Delta_0$, and
$\Delta_1$ can be found in Appendix \ref{appB}.

Equation (\ref{20})  holds for both $\lambda=\lambda_2$ and $\lambda=\lambda_1$.
These two possibilities turn out to differ in that $\lambda_1^{(0)}$ (the value of
$\lambda_1$ at $x_1=0$) is a root of $\Delta_0$.
It is then important to keep track of the finite $x_1$ correction to $\lambda_1^{(0)}$
that is present in $\lambda_1$. To first order in $x_1$,
we have
\begin{equation}
\label{29} \lambda_2(a^*,x_1)\approx \lambda_2^{(0)}(a^*)+\lambda_2^{(1)}(a^*) x_1
\end{equation}
and
\begin{equation}
\label{29.1}
\lambda_1(a^*,x_1)\approx \lambda_1^{(0)}(a^*)+\lambda_1^{(1)}(a^*) x_1,
\end{equation}
where $\lambda_2^{(0)}$ and $\lambda_1^{(0)}$ are given by Eqs.\ (\ref{16}) and (\ref{17}), respectively.
The expressions of $\lambda_2^{(1)}$ and $\lambda_1^{(1)}$ can be obtained from the
general sixth-degree polynomial equation (\ref{3.15}); they are given in Appendix \ref{appB}.
It can be checked that
if $\lambda=\lambda_2^{(0)}$ in Eq.\ (\ref{20}), then [according to Eqs.\ (\ref{b1}) and
(\ref{b2})] $D(\lambda_2^{(0)})\neq 0$, $\Delta_0(\lambda_2^{(0)})\neq 0$, and
so the energy ratio $E_1/E$ vanishes when $x_1\to 0$, as may have been expected. However, if
$\lambda=\lambda_1^{(0)}$ in Eq.\ (\ref{20}), $\Delta_0(\lambda_1^{(0)})=0$ which implies that
$E_1/E\neq 0$. Therefore, by taking the tracer limit in Eq.\ (\ref{20}), one gets
\begin{equation}
\label{30} \lim_{x_1\to
0}\frac{1}{x_1}p_1^*(\lambda_2(a^*,x_1),a^*,x_1)=\frac{D(\lambda_2^{(0)}(a^*),a^*)}
{\Delta_0(\lambda_2^{(0)}(a^*),a^*)},
\end{equation}
\begin{eqnarray}
\label{31} \lim_{x_1\to
0}p_1^*(\lambda_1(a^*,x_1),a^*,x_1)&=&\frac{D(\lambda_1^{(0)}(a^*),a^*)}
{\Delta_{01}(a^*)+\Delta_1(\lambda_1^{(0)}(a^*),a^*)} \nonumber\\
&\neq &  0,
%\equiv \Theta(a^*, \mu,\alpha_{11},\alpha_{22},\alpha_{12}),
\end{eqnarray}
where
\begin{equation}
\label{32} \Delta_{01}(a^*)\equiv \left(\frac{\partial \Delta_0(\lambda,a^*)}{\partial
\lambda}\right)_{\lambda=\lambda_1^{(0)}(a^*)}\lambda_1^{(1)}(a^*).
\end{equation}
Equation (\ref{30}), where $\lambda_2$ is the argument of
$p_1^*$, is relevant for the case $\lambda_2^{(0)}>\lambda_1^{(0)}$
while conversely, Eq. (\ref{31}) applies when
$\lambda_2^{(0)}<\lambda_1^{(0)}$.
In Eq.\ (\ref{31}), use has been made of the fact that
$\Delta_0(\lambda_1^{(0)},a^*)=0$.
%The last equality of that Equation
%is a definition of the energy ratio in the ordered phase.

In conclusion, if $\lambda_{\text{max}}=\lambda_2^{(0)}$, the right-hand side of
Eq.\ (\ref{30}) is  the temperature ratio $T_1/T$ and so, the energy ratio $E_1/E=0$. On the other
hand, if $\lambda_{\text{max}}=\lambda_1^{(0)}$, the temperature ratio diverges to
infinity and the energy ratio becomes {\em finite}. The change from one behavior to the other is akin to a non-equilibrium transition
between disordered ($E_1/E=0$, \emmanuel{finite temperature ratio}) and ordered
($E_1/E\neq 0$, \emmanuel{diverging temperature ratio}) phases \cite{MSG96}.
The task now boils down to identifying the regions of parameter space where
$\lambda_1^{(0)}=\lambda_2^{(0)}$, from which the domains of existence of the
ordered and disordered phases can be obtained.
It is important to note here that although our procedure leads to tracer limit
quantities that {\it a priori} depend on the intruder-intruder coefficient of restitution
$\alpha_{11}$, such a parameter turns out to disappear from the final
expressions, \emmanuel{which is intuitively expected}.
We could not analytically show this property (the computer package of
symbolic calculations (MATHEMATICA) used was not able to provide a manageable expression), that is nevertheless
a systematic numerical observation. We will henceforth drop $\alpha_{11}$
from the relevant parameters in the tracer limit.

\subsection{Absence of shear rate (homogeneous cooling state)}

Before analyzing the physical consequences of the above mathematical treatment in
the shear flow problem ($a\neq 0$), it is instructive to consider the particular case
of vanishing shear rates. This state is referred to as the  homogeneous cooling state (HCS).
Such a situation has been \vicente{analyzed in detail
in a recent work \cite{GT12}. Here, for the sake of completeness, we summarize the most
relevant results in the HCS.}

When $a^*=0$, the roots
$\lambda_2^{(0)}$ and $\lambda_1^{(0)}$ simply reduce  to
\begin{eqnarray}
\label{18.1}
&\lambda_2^{(0)}=-\displaystyle\frac{1-\alpha_{22}^2}{2d}, \\
\label{18.2}
&\lambda_1^{(0)}=-\displaystyle\frac{2}{d}\,\mu_{21}(1+\alpha_{12})
\left[1-\frac{\mu_{21}}{2}(1+\alpha_{12})\right].
\end{eqnarray}
It is apparent that, even in the HCS, there are two
different regimes of behavior depending if $\lambda_2^{(0)}$ is smaller or
larger than $\lambda_1^{(0)}$. Equating $\lambda_2^{(0)}$ and $\lambda_1^{(0)}$ leads
to two critical mass ratios
\begin{equation}
\label{eq:muhcs}
\mu_{\text{HCS}}^{(-)}=\frac{\alpha_{12}-\sqrt{\frac{1+\alpha_{22}^2}{2}}}
{1+\sqrt{\frac{1+\alpha_{22}^2}{2}}} \qquad \hbox{and} \qquad
\mu_{\text{HCS}}^{(+)}=\frac{\alpha_{12}+\sqrt{\frac{1+\alpha_{22}^2}{2}}}
{1-\sqrt{\frac{1+\alpha_{22}^2}{2}}},
\end{equation}
with $\mu_{\text{HCS}}^{(-)}<\mu_{\text{HCS}}^{(+)}$.
When $\mu_{\text{HCS}}^{(-)}<\mu<\mu_{\text{HCS}}^{(+)}$, we have $\lambda_2^{(0)}>\lambda_1^{(0)}$
and the temperature ratio $T_1/T$ remains finite (disordered phase). It is given by \cite{GT12}
\begin{equation}
\label{18.4}
\frac{T_1}{T}=\frac{2\mu_{12}\mu_{21}(1+\alpha_{12})^2}{4\mu_{21}(1+\alpha_{12})
\left[1-\frac{\mu_{21}}{2}(1+\alpha_{12})\right]-1+\alpha_{22}^2}.
\end{equation}
The expressions of $\mu_{\text{HCS}}^{(+)}$ and $T_1/T$ coincide with the ones previously derived by
Ben-Naim and Krapivsky \cite{NK02} in their analysis on the velocity statistics of an
impurity immersed in a uniform granular fluid.
When $\mu>\mu_{\text{HCS}}^{(+)}$ or $\mu<\mu_{\text{HCS}}^{(-)}$,
$\lambda_2^{(0)}<\lambda_1^{(0)}$, so that $T_1/T\to \infty$ {\em but} the energy ratio
$E_1/E$ is finite. This is a new result together with the identification of
the bound $\mu_{\text{HCS}}^{(-)}$. The explicit expression of $E_1/E$ is \cite{GT12}
\begin{equation}
\label{18.4.1}
\frac{E_1}{E} \,=\,
\frac{\alpha_{22}^2-1+4\mu_{21}(1+\alpha_{12})\left[1-\frac{\mu_{21}}{2}(1+\alpha_{12})\right]}
{\alpha_{22}^2-1+2\mu_{21}(1-\alpha_{12}^2)}.
\end{equation}
\begin{figure}
\includegraphics[width=0.9\columnwidth]{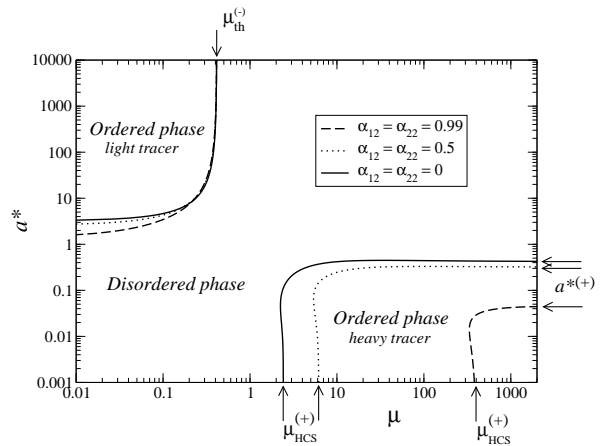}
\caption{Phase boundaries discriminating, in the tracer limit $x_1\to 0$ and for hard
spheres ($d=3$), between ordered and disordered
phases. In the ordered phases, the energy ratio $E_1/E$
(impurity over total energy) is finite --hence a diverging temperature ratio
$T_1/T$-- whereas in the disordered phase, the order parameter $E_1/E$ vanishes
with a corresponding finite temperature ratio.
The vertical arrows in the lower panel
indicate the threshold $\mu_{\text{HCS}}^{(+)}$ as given by
Eq. (\ref{eq:muhcs}). The values of $\mu_{\text{th}}^{(-)}=\sqrt{2}-1$
[here, common to all three parameter sets, see Eq. (\ref{19})]
and of $a^{*(+)}$ [given by Eq. (\ref{eq:astarplus})] are also indicated.
\label{fig:4}}
\end{figure}

Three remarks are in order here:
\begin{itemize}
\item while the upper bound $\mu_{\text{HCS}}^{(+)}$ is well defined for all values
of $\alpha_{12}$ and $\alpha_{22}$, the lower one is meaningful (i.e. positive)
 only when $\alpha_{12}>\sqrt{(1+\alpha_{22}^2)/2}$. Such a constraint cannot
be met when $\alpha_{12}=\alpha_{22}$ nor when
$\alpha_{12}<\sqrt{2}/2$, and requires ``asymmetric'' coefficients of restitution
(the above inequality implies $\alpha_{12}>\alpha_{22}$,
i.e. more dissipative inter host gas collisions than cross
intruder-gas encounters).
\item a similar extreme breakdown of the energy equipartition has been found \cite{SD01} for inelastic
hard spheres since, in the ordered phase ($T_1/T\to \infty$), the ratio of the mean square velocities for
the impurity and gas particles $m_2T_1/m_1T$ becomes \emph{finite} for an extremely heavy impurity particle
($m_1/m_2 \to \infty$). \emmanuel{Our Maxwell approach hence captures
robust effects, but at the same time exaggerates the trend of more refined models}.
\item in the present unforced situation, the results are independent of
the space dimension $d$.

\end{itemize}

\subsection{Nonzero shear rate}

We now wish to assess the influence of the shear rate on the transition observed in the
absence of shear (HCS). When $a^*=0$, we have seen that an ordered phase exists
for heavy impurities, while asymmetric dissipation may open a window for a
light impurity ordered phase, provided $\alpha_{12}>\sqrt{(1+\alpha_{22}^2)/2}$. On the other
hand, it is known for elastic systems that an ordered phase sets in for
$\mu < \sqrt{2} - 1$ \cite{note} (which can be seen as a
light impurity condition), provided the shear rate $a^*$ is larger than a certain
critical value  $a_c^*(\mu)$. To see how these two limiting cases are connected, we first show in Fig.
\ref{fig:4} how the shear rate affects the HCS scenario, and how collisional dissipation modifies the results obtained for ordinary gases ($\alpha_{rs}=1$) \cite{MSG96}.
Since we have chosen symmetric dissipation parameters
($\alpha_{12}=\alpha_{22}$) in Fig. \ref{fig:4}, the light tracer ordered phase is precluded
on the $a^*=0$ axis. It can be seen that this phase exists, in the shear rate versus
mass ratio plane, provided $a^*>a_c^*$ and
$\mu<\mu_{\text{th}}^{(-)}$, which defines
a domain that is rather insensitive to the value of the coefficient of restitution.
On the other hand,
the heavy tracer ordered phase is much more sensitive to collisional dissipation.
At variance with the light tracer phase,
it has an enhanced domain of existence for more dissipative systems,
and disappears in the elastic limit, as suggested by Fig.\ \ref{fig:4}
(see the dashed line corresponding to $\alpha_{12}=\alpha_{22}=0.99$,
squeezed in the lower right corner of the graph).
We also note that the heavy impurity ordered phase is destroyed by a sufficiently
vigorous shear rate, see below.

%\null\vskip 1ex
\begin{figure}[t]
\includegraphics[width=0.9\columnwidth]{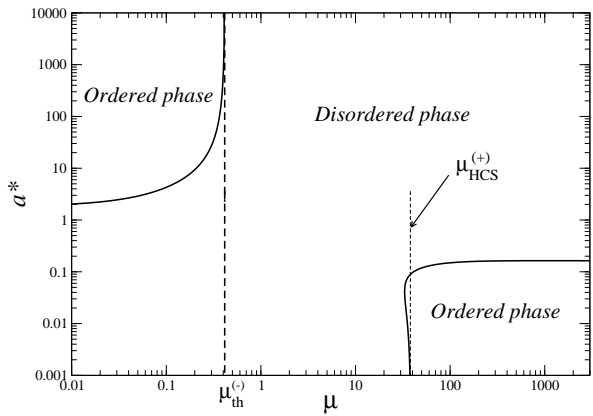}
\null\vskip 1ex
\caption{Same as Fig. \ref{fig:4}, for hard disks ($d=2$). Here,
$\alpha_{12}=\alpha_{22} =0.9$.
\label{fig:4bis}}
\end{figure}
%\null\vskip 1ex
\begin{figure}[b]
\includegraphics[width=0.9\columnwidth,clip]{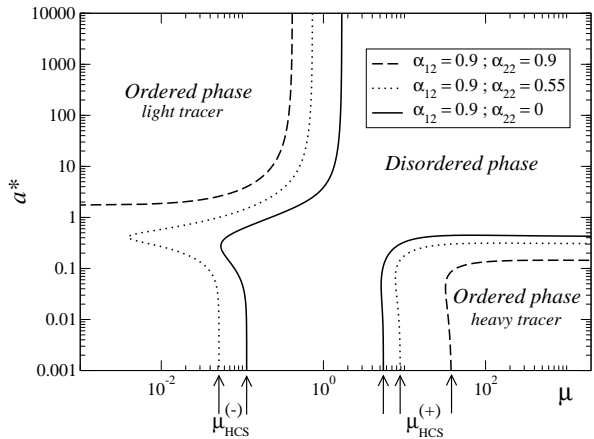}
\caption{Tracer limit phase diagram, in three dimensions.
The values of $\mu_{\text{HCS}}^{(-)}$ and $\mu_{\text{HCS}}^{(+)}$
predicted by
Eq. (\ref{eq:muhcs}) are shown by the arrows. For $\alpha_{12}=0.9$ as chosen here,
the constraint $\alpha_{12}>\sqrt{(1+\alpha_{22}^2)/2}$
for the existence of a light impurity ordering at vanishing
shear, reads $\alpha_{22}<0.787$. The dashed line therefore does
not reveal any light impurity ordered phase at zero shear rate
\emmanuel{(this was also the case in Fig. \ref{fig:4})}.
\label{fig:6}}
\end{figure}

The characteristic features seen in Figure \ref{fig:4} --with two distinct ordering
pockets-- may be rationalized by further
analytical results. In particular, the threshold $\mu_{\text{th}}^{(-)}$
may be obtained by the condition $\lambda_1^{(0)}=\lambda_2^{(0)}$ when $a^*\to \infty$. Following that route, we find
\begin{equation}
\label{19}
\mu_{\text{th}}^{(-)} \,=\, \sqrt{2}
\frac{1+\alpha_{12}}{1+\alpha_{22}}-1,
\end{equation}
that does not depend on space dimension, as the thresholds pertaining
to the HCS. As expected, we recover the
threshold value $\mu_{\text{th}}^{(-)}=\sqrt{2}-1$ for elastic systems.
Likewise, the upper shear rate $a^{*(+)}$ beyond which the heavy impurity ordered phase
disappears may be derived from enforcing the limit $\mu \to \infty$ in the equality
$\lambda_2^{(0)}=\lambda_1^{(0)}$. Since $\mu_{21}\to 0$, then $\lambda_1^{(0)}\to 0$ so that $\lambda_2^{(0)}=0$. This implies [see Eq.\ \eqref{16}]
\begin{equation}
\label{19.1}
F(\widetilde{a})=\frac{d+2}{2d}\frac{1-\alpha_{22}^2}{1+\alpha_{22}^2}.
\end{equation}
Equation \eqref{19.1} yields
\begin{equation}
a^{*(+)} \,=\, \frac{1+d-\alpha_{22}}{d} \,
\sqrt{\frac{1-\alpha_{22}^2}{2(d+2)}},
\label{eq:astarplus}
\end{equation}
where use has been made of the relation
%\begin{widetext}
\begin{equation}
\label{F}
 F (1 + 2F)^2=\frac{x^2}{d},
\end{equation}
%\end{widetext}
where $F(x)$ is given by Eq.\ \eqref{17.1}. The identity \eqref{F}
allows for a convenient expression of $\widetilde a$ [and hence $a^*$ through
Eq. (\ref{18})], once $\lambda$ [and hence $F(\widetilde a)$] is known.
Since the value of $a^{*(+)}$ has been obtained from the condition $\lambda_2^{(0)}=0$
(constant temperature \vicente{in the ordered phase}), the expression \eqref{eq:astarplus}
also gives the $\alpha$-dependence of the shear rate in the steady USF state \cite{GT10}.
%\null\vskip 1ex
\begin{figure}
\includegraphics[width=0.9\columnwidth,clip]{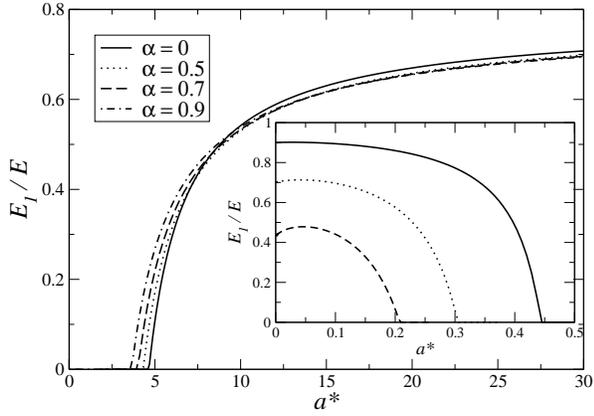}
\caption{Order parameter $E_1/E$ as a function of the reduced shear rate $a^*$,
for different values of $\alpha\equiv\alpha_{12}=\alpha_{22}$. The main graph
is for $\mu=0.1$ (light impurity). In the inset, where the results for $\mu=20$ are displayed
with the same convention for the different curves as in the main graph,
the value of $\alpha=0.9$ is not shown, since with such a
parameter, $\mu_{\text{HCS}}^{(+)}>\mu$, hence no ordering is possible.
In other words, the counterpart of the dashed-dotted line of the main graph
is simply $E_1/E=0$ in the inset.
\label{fig:7}}
\end{figure}

The quantities $\mu_{\text{th}}^{(-)}$ and $a^{*(+)}$ are indicated
by arrows in Fig. \ref{fig:4}. Consistent with the disappearance of
the heavy impurity ordered phase for elastic systems, is the vanishing
of $a^{*(+)}$ when $\alpha_{22}\to 1$. It can also be noted that $a^{*(+)}$
is only a host property, and does not depend on $\alpha_{12}$.
Finally, Fig. \ref{fig:4bis} shows that the behavior in two and
three dimensions are similar.

We now turn to asymmetric collisional dissipation cases, so that the light impurity
ordered phase may exist even for vanishing shear rates. Such a scenario is
illustrated in Fig. \ref{fig:6}, that corroborates the
analytical predictions. The boundary of the light impurity ordered phase
in a shear-mass ratio phase diagram is non trivial,
and indicates the existence of
an interval of $\mu$ values, below $\mu_{\text{HCS}}^{(-)}$,
with a re-entrance feature. Indeed, starting from the ordered phase at
$a^*=0$, and increasing $a^*$ at fixed $\mu$, one first meets a transition
from order to disorder, followed by a subsequent ordering
transition. Similarly, and again for $\alpha_{12}>\sqrt{(1+\alpha_{22}^2)/2}$,
the following series order $\to$ disorder $\to$ order occurs
when $\mu$ is increased at fixed reduced shear rate, provided
$a^*<a^{*(+)}$.
%\null\vskip 1ex
\begin{figure}[h]
\includegraphics[width=0.9\columnwidth]{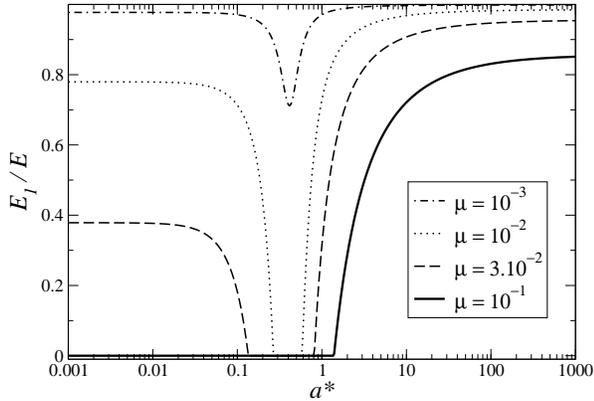}
\caption{Order parameter for $d=3$, $\alpha_{12} = 0.9$ and $\alpha_{22}=0.55$
(see the dotted line in Fig. \ref{fig:6}). The value $\mu=10^{-1}$
leads to a disordered phase at zero shear rate (thick continuous curve),
while the smaller
values of $\mu$ reported here are associated with small shear ordering,
with the possibility of an intermediate disordered phase (see text).
\label{fig:8}}
\end{figure}

\begin{figure}
\includegraphics[width=0.9\columnwidth,clip]{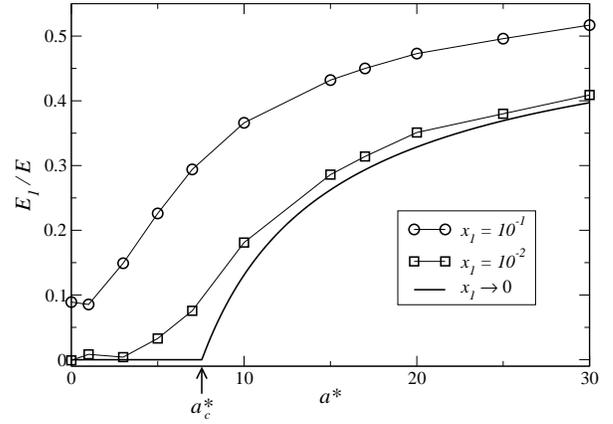}
\caption{Shear rate dependence of $E_1/E$ for hard spheres ($d=3$) in the case
$\mu=0.2$ and $\alpha_{22}=\alpha_{12}=0.9$. Three different values of the mole
fraction $x_1$ have been considered: $x_1=0.1$ (circles), $x_1=0.01$
(squares), and $x_1=0$ (solid line). }\label{Fig:9}
\end{figure}

To substantiate the phase diagram reported above, we show in Fig.
\ref{fig:7} the shear rate dependence of the order parameter $E_1/E$,
for different values of the (common) coefficient of restitution
and mass ratios. The light impurity ordering
is seen to be enhanced by increasing the shear rate,
while the reverse behavior is in general observed for
heavy impurities (see the inset). The critical thresholds observed in Fig.
\ref{fig:7} are fully compatible with those appearing in Fig. \ref{fig:4}.
Focussing next on the light impurity ordered phase, we report
the order parameter variation  in cases of asymmetric collisional dissipation.
To this end, we return to the set $\alpha_{12}=0.9$ and $\alpha_{22}=0.55$
addressed in Fig. \ref{fig:7}. For such quantities, one has
$\mu_{\text{HCS}}^{(-)} \simeq 0.051$. As can be seen in Fig.
\ref{fig:8}, when $\mu<\mu_{\text{HCS}}^{(-)}$, the order
parameter is non vanishing at small shear rates, while when
$\mu_{\text{HCS}}^{(-)} <\mu < \mu_{\text{th}}^{(-)}$,
ordering sets in only beyond a critical shear rate
(note that $\mu_{\text{th}}^{(-)} \simeq 0.733$, so that
the light impurity ordered phase does exist in some
shear domain, for all the mass ratios used in Fig. \ref{fig:8}).
The figure clearly shows the re-entrance of order alluded to above
(see the curves for $\mu=10^{-2}$ and $\mu=3\times 10^{-2}$, where an
intermediate interval of $a^*$ values leads to a disordered phase with
$E_1/E=0$. For $\mu=10^{-3}$ (dashed-dotted line), all shear rates
lead to phase ordering ($E_1/E \neq 0$), but there is a fingerprint of
the reentrant behavior in the non monotonicity of
the order parameter with $a^*$.

For completeness, we now show how the tracer limit is approached.
For $\mu < \mu_{\text{th}}^{(-)}$ and $\alpha_{12}=\alpha_{22}$,
ordering occurs when the system is driven sufficiently far from equilibrium
($a^*>a_c^*$). To illustrate how the abrupt transition observed
in the tracer limit is blurred by finite concentration,
Fig.\ \ref{Fig:9} shows $E_1/E$ versus the (reduced) shear rate $a^*$ for $d=3$,
$m_1/m_2=0.2$, $\alpha_{11}=\alpha_{22}=\alpha_{12}=0.9$ and $x_1=10^{-1}$, $10^{-2}$
and $0$. It is apparent that the curves tend to collapse
to the exact tracer limit result
as the mole fraction $x_1$ vanishes. This is indicative of the
consistency of the analytical
results derived at $x_1=0$. Moreover, since the impurity particle is sufficiently
lighter than
the particles of the gas ($\mu=0.2<\mu_{\text{th}}^{(-)}\simeq 0.414$), the energy ratio
$E_1/E$ is non vanishing in
the tracer limit if $a^*>a_c^*\simeq 7.557$ in the present case (see the figure).
It can also be noted that at finite $x_1$, the energy ratio for small shear rates
is of the order of $x_1$, since the temperature ratio is quite close
to unity in that limit (see e.g. Fig. \ref{Fig:2}).
\begin{figure}[h]
\includegraphics[width=0.9\columnwidth]{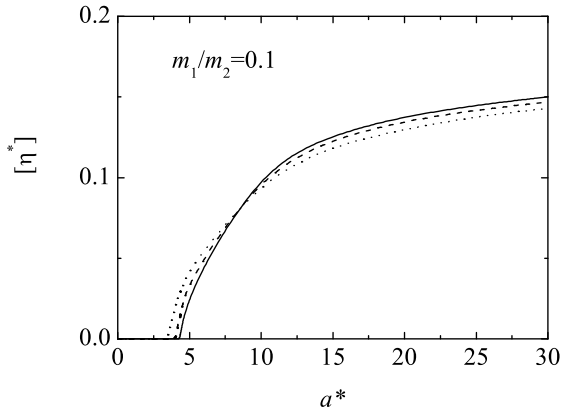}
\caption{Shear rate dependence of the intrinsic viscosity $[\eta^*]$ for hard spheres ($d=3$) in the case
$m_1/m_2=0.1$. Three different values of the (common) coefficient of restitution $\alpha\equiv\alpha_{12}=\alpha_{22}$ have been considered: $\alpha=0.5$ (solid line), $\alpha=0.7$ (dashed line), and $\alpha=1$ (dotted line).
\label{fig11}}
\end{figure}

\vicente{Although we have focused our attention on the energy ratio,
it is clear that similar features can be analyzed when considering other quantities.
An interesting candidate is the \emph{non-linear} shear viscosity $\eta^*$ defined as
\begin{equation}
\label{a12}
\eta^*(a^*)=-\frac{P_{xy}^*}{a^*},
\end{equation}
where the shear stress $P_{xy}^*=P_{1,xy}^*+P_{2,xy}^*$. The rheological function $\eta^*$ characterizes the nonlinear response of the system to the action of strong shearing. In terms of $p_1^*$ and for finite values of $x_1$, the expressions of $P_{1,xy}^*$ and $P_{2,xy}^*$ are given by Eqs.\ \eqref{a10} and \eqref{a11}, respectively.

As expected, in the tracer limit ($x_1\to 0$) and in the disordered phase, the total shear viscosity $\eta^*$ of the mixture coincides with that of the solvent gas $\eta^*\to \eta_s^*$, where $\eta_s^*$ is given by Eq.\ \eqref{a20}. However, in the ordered phase, there is a \emph{finite} contribution to the total shear viscosity coming from the tracer particles (see Eq.\ \eqref{a21}). To illustrate it, Fig.\ \ref{fig11} shows the shear rate dependence of the \emph{intrinsic} shear viscosity \cite{Y71} $[\eta^*]=\lim_{x_1\to 0}(\eta^*-\eta_s^*)/\eta_s^*$ for the mass ratio $m_1/m_2=0.1$ (light impurity) and different values of the (common) coefficient of restitution. We observe that the intrinsic viscosity $[\eta^*]$ is clearly different from zero for shear rates larger than its corresponding critical value. On the other hand, its magnitude is smaller than the one obtained for the order parameter $E_1/E$ (see Fig.\ \ref{fig:7})}

%%%%%%%%%%%%%%%%%%%%%%%%%%%%%%%%%%%%%%%%%%%%%%%%%
\section{Discussion and conclusion}
\label{sec5}

In this paper we have analyzed the dynamics of an impurity immersed in a granular gas subject to
USF.
The study has been performed in two successive steps. First, the pressure tensor of a granular
binary mixture of inelastic Maxwell gases under USF has been obtained from an \emph{exact} solution of
the Boltzmann equation. This solution applies for arbitrary values of the shear rate $a$ and the
parameters of the mixture, namely, the mole fraction $x_1=n_1/n$, the mass ratio $\mu\equiv m_1/m_2$
and the coefficients of restitution $\alpha_{11}$, $\alpha_{22}$ and $\alpha_{12}$. Then, the tracer
limit ($x_1\to 0$) of the above solution has been carefully considered, showing that the relative
contribution of the tracer species to the total properties of the mixture does {\em not} necessarily vanish
as $x_1\to 0$. This surprising result extends
to \emph{inelastic} gases some results derived some time ago for ordinary gases \cite{MSG96}.
The above phenomenon can be seen as a non-equilibrium phase transition, where the
relative contribution of the impurity to the total kinetic energy
$E_1/E$ plays the role of an order parameter \cite{rque20}.

The transition problem addressed here has been analyzed in the framework of the Boltzmann equation
with Maxwell kernel. The key advantage of inelastic Maxwell models,
in comparison with the more realistic inelastic hard
sphere model, is that the collisional moments of the Boltzmann collision operators $J_{rs}[f_r,f_s]$
can be exactly evaluated in terms of the velocity moments of $f_r$ and $f_s$, without the explicit
knowledge of these velocity
distribution functions. Here, we have explicitly determined the collisional moments associated with
the second-degree velocity moments to get the pressure tensor of the mixture. In addition, the collision rates
$\omega_{rs}$ appearing in the operators $J_{rs}[f_r,f_s]$ have been chosen to be time independent
so that the interaction model allows to disentangle the effects of collisional dissipation (accounted
for by the coefficients of restitution) from those of boundary conditions (embodied in the reduced shear
rate $a^*$ defined as $a^*=a/\nu_0$, $\nu_0$ being a characteristic collision frequency). Consequently,
within this model, collisional dissipation and viscous heating generally do not compensate, so that the
granular temperature increases (decreases) with time if viscous heating is larger (smaller) than
collisional cooling.

In our system, the temperature $T_2$ of the gas particles ($T_2\simeq T$
in the disordered phase when $x_1\to 0$) changes in
time due to two competing effects: the viscous heating term ($-aP_{xy}$)
and the inelastic cooling term ($\zeta T_2$). In fact, for long times,
$T_2(t)\sim \exp[\lambda_2^{(0)}\nu_0 t]$ where
$\lambda_2^{(0)}=-\zeta^*-(2a^*/d)P_{xy}^*$ (here, $\zeta^*=\zeta/\nu_0$ and
$P_{xy}^*=P_{xy}/p$). Since $P_{xy}^*<0$, the cooling rate $\zeta^*$ can be
interpreted as the ''thermostat'' parameter needed to get a stationary value
for the temperature $T_2$ of the gas particles. At a given value of the
shear rate, the cooling rate increases with dissipation and so, $T_2$
decreases as $\alpha_{rs}$ decreases. The tracer particles are also
subject to two antagonistic mechanisms. On the one hand, $T_1\to \infty$ due
to viscous heating and on the other hand, collisions with the gas particles
tend to ``thermalize'' $T_1$ to $T_2$. Both effects are accounted for by the
root $\lambda_1^{(0)}$ that governs the behavior of $T_1$ for long times,
i.e., $T_1(t)\sim \exp[\lambda_1^{(0)}\nu_0 t]$. When
$\lambda_1^{(0)}>\lambda_2^{(0)}$, the temperature ratio $T_1/T_2$ grows
without bounds. The parameter ranges where such a requirement is met
define the ordered ``pockets'' of the phase diagram, and where
the energy ratio $E_1/E$ --explicitly worked out here--
reaches a finite value.
We have found that two different families of ordered phase can be
encountered
\begin{itemize}
\item a light impurity phase, provided that the mass ratio
$\mu$ does not exceed the threshold $\mu_{\text{th}}^{(-)}$
given by Eq. (\ref{19}). Such a phase always exists for shear rates larger than
a certain critical value, but can also be observed at vanishing shear
in cases of asymmetric collisional dissipation, whenever gas-gas collisions are
sufficiently more dissipative that intruder-gas collisions
($\alpha_{12}> \sqrt{(1+\alpha_{22}^2)/2}$).
\item a heavy impurity phase, that
--unlike the light impurity phase--
cannot accommodate large shear, and requires $a^*<a^{*(+)}$,
where the threshold $a^{*(+)}$ is given by Eq. (\ref{eq:astarplus}).
\end{itemize}

\begin{figure}
\includegraphics[width=0.9\columnwidth]{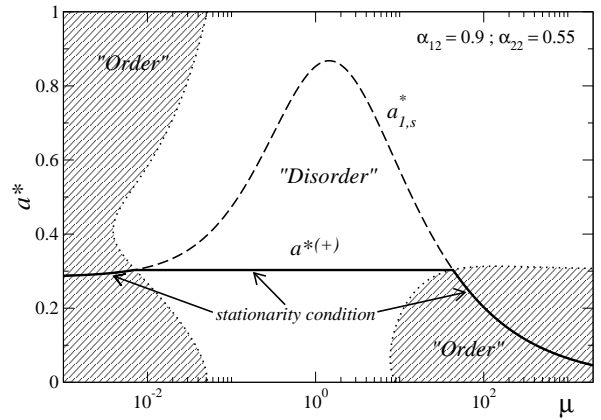}
\caption{Same plot as Fig. \ref{fig:6}, where only the curves pertaining to the
values $\alpha_{12}=0.9$ and $\alpha_{22}=0.55$ have been retained.
The ordered pockets are shown by the hatched areas.
The steady state condition (\ref{3.11}) of equal viscous heating and collisional dissipation,
implies a coupling between $a^*$ and $\mu$, that corresponds to the thick continuous line.
The bell shaped dashed curve shows the locus of points where $\lambda_1^{(0)}=0$.
It therefore coincides with the thick continuous line in the ordered phase.
However, it does not yield the steady state condition in the disordered case, where
the reduced shear rate is given by $a_{2,s}^*=a^{*(+)}$, that does not depend
on $\mu$ (hence the horizontal sector in the thick continuous line).
}
\label{Fig:11}
\end{figure}

The fact that $a^*$ and $\alpha_{rs}$ are independent parameters [unless $a^*$
takes the specific value $a_s^*$ given by the steady state condition (\ref{3.11})] allows one to
carry out a clean analytical study of the combined effect of both control parameters on the
properties of the impurity particle. It is however important to bring to the fore
the precise coupling between shear and collisional dissipation that the steady state
condition Eq. (\ref{3.11}) implies. The answer depends on the phase considered,
ordered or disordered and the energy balance embodied in Eq (\ref{3.11}) can be expediently expressed
as max$(\lambda_1^{(0)},\lambda_2^{(0)})=0$.  In other words, $a_s^*$ follows from
enforcing $\lambda_2^{(0)}=0$ in the disordered state, and likewise $\lambda_1^{(0)}=0$
in the ordered regime. The solution $a_{2,s}^*$ of the first equation
has already been displayed in Eq. (\ref{eq:astarplus})~:  $a_{2,s}^*=a^{*(+)}$.
On the other hand, the solution $a_{1,s}^*$ to
$\lambda_1^{(0)}=0$ can be obtained along similar lines and reads
\begin{equation}
a_{1,s}^*=\beta_{12}^{3/2}\sqrt{\frac{1-\frac{1}{2}
\beta_{12}}{d+2}}\left[1+\frac{d+2}{d}\left(\frac{2}{\beta_{12}}-1\right)\right]
\end{equation}
where we have introduced the notation $\beta_{12}\equiv \mu_{21}(1+\alpha_{12})$.
Finally, the steady state condition implies $a^* = a^{*(+)}$
(resp. $a^*= a_{1,s}^*$) in the disordered (resp ordered) case. The corresponding
line in the shear versus mass ratio plane is shown in Fig. \ref{Fig:11},
for a given set of dissipation parameters that corresponds to one of the
situations analyzed in Fig. \ref{fig:6}. It appears that remaining
on the steady state line shown by the thick curve,
one spans the three possible regimes: light tracer ordering at small
$\mu$, disordered phase at intermediate values, and heavy tracer
order at larger $\mu$. We conclude here that the scenario uncovered
in our analysis is not an artifact of having decoupled shear from dissipation,
and we emphasize that from a practical point of view, studying
the transient regime (before the steady state occurs) anyway offers the
possibility to enforce the above decoupling.

\begin{figure}
\includegraphics[width=0.9\columnwidth]{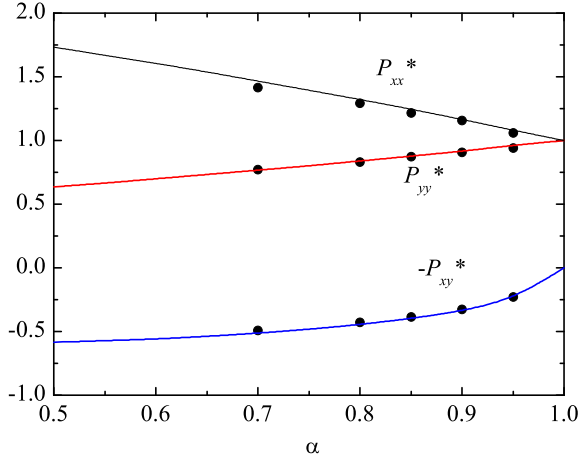}
\caption{(color online) \vicente{Plot of the reduced elements of the pressure tensor as functions of the (common) coefficient of restitution
$\alpha=\alpha_{11}=\alpha_{12}=\alpha_{22}$ for a three dimensional system ($d=3$) in the steady USF. The predictions of the inelastic Maxwell model (solid line,
present work)
are tested against Monte Carlo simulation data (symbols, taken
from Ref. \cite{MG02}). The parameters are $x_1=0.5$ (equimolar mixture) and $\mu=2$. It should be noted that for a meaningful
comparison, $a^*$ and $\alpha$ need to be coupled (see Eq.\ \eqref{3.11}), since inelastic hard spheres enjoy
a steady state for a precise value of the reduced shear rate,
that depends on the coefficient of restitution $\alpha$.}
\label{Fig:12}}
\end{figure}

In addition, we would like to stress here that, in spite of the
approximate nature of our plain vanilla Maxwell model,
the results obtained for binary mixtures under steady
USF \cite{G03} compare quite well with Monte Carlo simulations of inelastic hard spheres \cite{MG02}.
\vicente{This can be seen in Fig. \ref{Fig:12}  for the (reduced) elements of the pressure tensor $P_{ij}^*$ in the steady shear flow state defined by the condition \eqref{3.11}.}
Therefore, we expect that the transition found in this paper is not artefactual and can be detected
%(at least in the transient regime before the asymptotic steady state)
in the case of hard spheres interaction.

It is quite natural --when analyzing the dynamics of an impurity immersed in a
background of mechanically different particles-- to invoke two assumptions.
First, that the state of the solvent (excess component 2) is
not affected by the presence of the tracer (solute) particles 1. Second, that
the effect on the state of the solute due to collisions among the tracer
particles themselves can be neglected. We have seen that the
second expectation is correct (the coefficient of restitution $\alpha_{11}$
is immaterial in the tracer limit, a property that is not obvious
from the cumbersome analytical formulas reported here), but
the first expectation is invalidated in the regions where the ordered phase sets in.
Consequently, the seemingly natural ``Boltzmann-Lorentz'' point of view
--with the
one-particle velocity distribution function $f_2$ of the granular gas
obeying a (closed) nonlinear Boltzmann kinetic equation while the one-particle
velocity distribution function $f_1$ of the impurity particle obeys a linear
Boltzmann-Lorentz kinetic equation-- breaks down.
Our results show that collisions
of type 2-1 affect $f_2$, despite
being much less frequent than collisions of type 2-2.
We conclude here that, rather unexpectedly, the
tracer problem is as complex as the general case
of a binary mixture at arbitrary mole fractions.

\acknowledgments

The research of V. G. has been supported by the Ministerio de Educaci\'on y Ciencia
(Spain) through Grant No. FIS2010-16587, partially financed by FEDER funds and by
the Junta de Extremadura (Spain) through Grant No. GR10158.

%%%%%%%%%%%%%%%%%%%%%%%%%%%%%%%%%%%%%%%%%%%%%%%%%%%%%%%%%%%%%%%%%%%%%%%%%%%%%%%%%%%%%%%%%%%%%%%%%%%%%%%%%%%%
\appendix
\section{Energy ratio $p_1^*$ and shear stress $P_{xy}^*$
\label{appA}}
The expression of the energy ratio $p_1^*$ can be obtained from Eq.\ (\ref{7}). It can be written as
\begin{equation}
\label{a5}
p_1^*=\frac{K a^{*2}+L}{R a^{*2}+S},
\end{equation}
where
\begin{eqnarray}
\label{a6}
K&=&-2A_{12}^*\lambda^2+4(A_{22}^*B_{12}^*-A_{12}^*B_{22}^*)\lambda+2A_{22}^*
B_{12}^*\nonumber\\
& & \times (B_{11}^*+B_{22}^*)
-2A_{12}^*(B_{12}^*B_{21}^*+B_{22}^{*2}),
\end{eqnarray}
\begin{equation}
\label{a7}
L=d(B_{12}^*-A_{12}^*)\left[\lambda^2+(B_{11}^*+B_{22}^*)\lambda+B_{11}^*B_{22}^*-
B_{12}^*B_{21}^*\right]^2,
\end{equation}
\begin{eqnarray}
\label{a8}
R&=&2(A_{11}^*-A_{12}^*)\lambda^2-4\left[B_{12}^*(A_{21}^*-A_{22}^*)\right.\nonumber\\
& & \left. +B_{22}^*(A_{12}^*
-A_{11}^*)\right]\lambda+2B_{12}^*(B_{11}^*+B_{22}^*) \nonumber\\
& & \times (A_{22}^*-A_{21}^*)+2(A_{11}^*-A_{12}^*)
(B_{12}^*B_{21}^*+B_{22}^{*2}),\nonumber\\
\end{eqnarray}
\begin{eqnarray}
\label{a9}
S&=&d(A_{11}^*-A_{12}^*-B_{11}^*+B_{12}^*-\lambda)\nonumber\\
& & \times \left[\lambda^2+(B_{11}^*+B_{22}^*)\lambda+
B_{11}^*B_{22}^*-B_{12}^*B_{21}^*\right]^2.\nonumber\\
\end{eqnarray}

\vicente{Once the energy ratio is known, the remaining relevant elements of the pressure tensor can be easily obtained from Eqs.\ \eqref{7}--\eqref{9.1}. In particular, the shear stress $P_{xy}^*$ is given by  $P_{xy}^*=P_{1,xy}^*+P_{2,xy}^*$ where
\begin{equation}
\label{a10}
P_{1,xy}^*=\frac{d\left[A_{12}^*-B_{12}^*-\left(B_{11}^*+\lambda-A_{11}^*+
A_{12}^*-B_{12}^*\right)p_1^*\right]}{2a^*},
\end{equation}
\begin{equation}
\label{a11}
P_{2,xy}^*=\frac{d\left[A_{21}^*-B_{21}^*-\left(B_{22}^*+\lambda-A_{22}^*+
A_{21}^*-B_{21}^*\right)(1-p_1^*)\right]}{2a^*}.
\end{equation}
}

%%%%%%%%%%%%%%%%%%%%%%%%%%%
\section{Derivation of $\lambda_1^{(0)}$ and $\lambda_2^{(0)}$}
\label{appAbis}

When $x_1\to 0$, the sixth-degree
equation (\ref{3.15}) for the rates $\lambda$ factorizes into two cubic equations given by
\begin{equation}
\label{10}
2 a^{*2}A_{22}^{(0)}+d(A_{22}^{(0)}-B_{22}^{(0)}-\lambda)(B_{22}^{(0)}+\lambda)^2=0,
\end{equation}
\begin{equation}
\label{11}
2 a^{*2}A_{11}^{(0)}+d(A_{11}^{(0)}-B_{11}^{(0)}-\lambda)(B_{11}^{(0)}+\lambda)^2=0,
\end{equation}
where $A_{rs}^{(0)}$ and $B_{rs}^{(0)}$ denote the zeroth-order contributions to the expansion of
$A_{rs}^*\equiv A_{rs}/\nu_0$ and $B_{rs}^*\equiv B_{rs}/\nu_0$, respectively, in powers of $x_1$.
They are given by
\begin{equation}
\label{12}
A_{22}^{(0)}=\frac{(1+\alpha_{22})^2}{2(d+2)},
\end{equation}
\begin{equation}
\label{13}
B_{22}^{(0)}=\frac{(1+\alpha_{22})(d+1-\alpha_{22})}{d(d+2)},
\end{equation}
\begin{equation}
\label{14}
A_{11}^{(0)}=\frac{\mu_{21}^2}{d+2}(1+\alpha_{12})^2,
\end{equation}
\begin{equation}
\label{15}
B_{11}^{(0)}=\frac{2}{d(d+2)}\mu_{21}(1+\alpha_{12})\left[d+2-\mu_{21}(1+\alpha_{12})\right].
\end{equation}
Equation (\ref{10}) is associated with the time
evolution of the excess component. Its largest root is given by Eq. (\ref{16}).
On the other hand,
Eq. (\ref{11}) gives the transient behavior of the impurity. Its largest root is
given by Eq. (\ref{17}).

%%%%%%%%%%%%%%%%%%%%%%%%%%%
\section{Some explicit expressions in the tracer limit}
\label{appB}
\begin{widetext}
In this Appendix, we provide some of the expressions used along the text in the tracer limit. First, the quantities $D$, $\Delta_0$ and $\Delta_1$ appearing in Eq.\ (\ref{20}) are given by
\begin{equation}
\label{b1}
D(\lambda)=d(B_{12}^{(1)}-A_{12}^{(1)})(B_{11}^{(0)}+\lambda)^2(B_{22}^{(0)}+\lambda)^2
+2a^{*2}\left[
A_{22}^{0)}B_{12}^{(1)}(B_{11}^{(0)}+B_{22}^{(0)}+2\lambda)-A_{12}^{(1)}(B_{22}^{(0)}+\lambda)^2\right],
\end{equation}
\begin{equation}
\label{b2}
\Delta_0(\lambda)=(B_{22}^{(0)}+\lambda)^2\left[ 2a^{*2}A_{11}^{(0)}+d(A_{11}^{(0)}-B_{11}^{(0)}-
\lambda)(B_{11}^{(0)}+\lambda)^2\right],
\end{equation}
%\begin{widetext}
\begin{eqnarray}
\label{b3}
\Delta_1&=&d(B_{11}^{(0)}+\lambda)(B_{22}^{(0)}+\lambda)\left\{(A_{11}^{(1)}-A_{12}^{(1)}-B_{11}^{(1)}
+B_{12}^{(1)})\right. \nonumber\\
& &  \times(B_{11}^{(0)}+\lambda)(B_{22}^{(0)}+\lambda)+2(A_{11}^{(0)}-B_{11}^{(0)}-\lambda)\nonumber\\
& & \times \left.\left[
B_{22}^{(1)}(B_{11}^{(0)}+\lambda)+B_{11}^{(1)}(B_{22}^{(0)}+\lambda)-
B_{21}^{(0)}B_{12}^{(1)}\right]\right\}\nonumber\\
& &
+2a^{*2}\left\{(A_{22}^{(0)}-A_{21}^{(0)})B_{12}^{(1)}(B_{11}^{(0)}+B_{22}^{(0)}+2\lambda)\right.\nonumber\\
& & \left.+
(B_{22}^{(0)}+\lambda)\left[
(B_{22}^{(0)}+\lambda)(A_{11}^{(1)}-A_{12}^{(1)})+2A_{11}^{(0)}B_{22}^{(1)}\right]
+A_{11}^{(0)}B_{21}^{(0)}B_{12}^{(1)}\right\}.
\end{eqnarray}
%\end{widetext}
In these equations, $A_{rr}^{(0)}$ and $B_{rr}^{(0)}$ are given by Eqs.\ (\ref{12})-(\ref{15}) and
\begin{equation}
\label{b4} A_{21}^{(0)}=A_{12}^{(1)}=\frac{\mu_{12}\mu_{21}}{(d+2)}(1+\alpha_{12})^2,
\quad B_{21}^{(0)}=B_{12}^{(1)}=-\frac{2}{d}A_{21}^{(0)},
\end{equation}
\begin{equation}
\label{b5}
A_{11}^{(1)}=\frac{1}{2(d+2)}(1+\alpha_{11})^2,
\end{equation}
\begin{equation}
\label{b6} B_{11}^{(1)}=\frac{1}{d(d+2)}(1+\alpha_{11}) (d+1-\alpha_{11}),
\end{equation}
\begin{equation}
\label{b7}
B_{22}^{(1)}=\frac{2}{d(d+2)}\mu_{12}(1+\alpha_{12})\left[d+2-\mu_{12}(1+\alpha_{12})\right].
\end{equation}

The expressions of the coefficients $\lambda_i^{(1)}(a)$ can be written as
\begin{equation}
\label{an10}
\lambda_i^{(1)}=\frac{X_i(\lambda_i^{(0)})}{Y_i(\lambda_i^{(0)})},
\end{equation}
where
%\begin{widetext}
\begin{equation}
\label{an11}
X_i(\lambda_i^{(0)})=4 X_i^{(4)} a^{*4}+2 d X_i^{(2)} a^{*2}
+
d^2 (B_{11}^{(0)}+\lambda_i^{(0)})(B_{22}^{(0)}+\lambda_i^{(0)})X_i^{(0)},
\end{equation}
\begin{eqnarray}
\label{an12}
Y_i(\lambda_i^{(0)})&=&d (B_{22}^{(0)}+\lambda_i^{(0)})\left[
2 a^{*2}A_{11}^{(0)}d(A_{11}^{(0)}-B_{11}^{(0)}
-\lambda_i^{(0)})(B_{11}^{(0)}+
\lambda_i^{(0)})^2\right]\left[2 A_{22}^{(0)}-3(B_{22}^{(0)}+\lambda_i^{(0)})\right]\nonumber\\
& +& d (B_{11}^{(0)}+\lambda_i^{(0)})\left[
2 a^{*2}A_{22}^{(0)}+d(A_{22}^{(0)}-B_{22}^{(0)}-\lambda_i^{(0)})(B_{22}^{(0)}+\lambda_i^{(0)})^2\right]
\left[2 A_{11}^{(0)}-3(B_{11}^{(0)}+\lambda_i^{(0)})\right].
\end{eqnarray}
%\end{widetext}
Here, we have introduced the quantities
\begin{equation}
\label{an13}
X_i^{(4)}=A_{12}^{(1)}A_{21}^{(0)}-A_{11}^{(1)}A_{22}^{(0)}-A_{22}^{(1)}A_{11}^{(0)},
\end{equation}
\begin{eqnarray}
\label{an14}
X_i^{(2)}&=&B_{11}^{(0)}\left[B_{11}^{(0)}\left(A_{22}^{(1)}B_{11}^{(0)}+
3B_{11}^{(1)}A_{22}^{(0)}
-B_{12}^{(1)}A_{21}^{(0)}\right)+2B_{12}^{(1)2}A_{22}^{(0)}\right]\nonumber\\
& & -B_{12}^{(1)}B_{22}^{(0)}\left[A_{21}^{(0)}\left(B_{22}^{(0)}+B_{11}^{(0)}\right)-
A_{22}^{(0)}B_{21}^{(0)}\right]+A_{11}^{(1)}\left[B_{22}^{(0)3}-A_{22}^{(0)}\left(B_{22}^{(0)2}+
B_{11}^{(0)2}\right)\right]\nonumber\\
& &+A_{11}^{(0)}\left[B_{12}^{(1)2}\left(B_{11}^{(0)}+2B_{22}^{(0)}\right)-
A_{22}^{(1)}\left(B_{11}^{(0)2}+B_{22}^{(0)2}\right)+
3B_{22}^{(0)2}B_{22}^{(1)}\right.\nonumber\\
& & \left.
-2A_{22}^{(0)}
\left(B_{11}^{(1)}B_{11}^{(0)}+B_{12}^{(1)2}+B_{22}^{(1)}B_{22}^{(0)}\right)
\right]\nonumber\\
& & -A_{12}^{(1)}\left\{B_{21}^{(0)}\left[(B_{11}^{(0)}+B_{22}^{(0)})(B_{11}^{(0)}+3\lambda_i^{(0)})
+B_{22}^{(0)2}+3\lambda_i^{(0)2}\right]\right.\nonumber\\
& & \left.-A_{21}^{(0)}\left[
B_{11}^{(0)}(B_{11}^{(0)}+2\lambda_i^{(0)})
+B_{22}^{(0)2}+2\lambda_i^{(0)}(B_{11}^{(0)}+\lambda_i^{(0)})\right]\right\}\nonumber\\
& & +\lambda_i^{(0)}\left\{3B_{11}^{(0)}\left[A_{22}^{(1)}(B_{11}^{(0)}+\lambda_i^{(0)})+
2 A_{22}^{(0)}B_{11}^{(1)}-B_{12}^{(1)}A_{21}^{(0)}\right]\right.
\nonumber\\
& &
-3B_{12}^{(1)}
\left[A_{21}^{(0)}(B_{22}^{(0)}+\lambda_i^{(0)})-
 A_{22}^{(0)}B_{21}^{(0)}\right]
+3A_{22}^{(0)}B_{11}^{(1)}\lambda_i^{(0)}+
 A_{22}^{(1)}\lambda_i^{(0)2}\nonumber\\
& &
+A_{11}^{(1)}\left[3B_{22}^{(0)}(B_{22}^{(0)}+\lambda_i^{(0)})+\lambda_i^{(0)2}-2A_{22}^{(0)}
(B_{11}^{(0)}+B_{22}^{(0)}+\lambda_i^{(0)})\right]\nonumber\\
& & \left.+A_{11}^{(0)}\left[3B_{21}^{(0)}B_{12}^{(1)}-2A_{22}^{(0)}(B_{11}^{(1)}+B_{22}^{(1)})
-2A_{22}^{(1)}(B_{11}^{(0)}+B_{22}^{(0)}+\lambda_i^{(0)})
+3B_{22}^{(1)}
(2B_{22}^{(0)}+\lambda_i^{(0)})\right]\right\},\nonumber\\
\end{eqnarray}
\begin{eqnarray}
\label{an15}
X_i^{(0)}&=&2(A_{11}^{(0)}-B_{11}^{(0)}-\lambda_i^{(0)})(A_{22}^{(0)}-B_{22}^{(0)}-\lambda_i^{(0)})
\left[B_{21}^{(0)}B_{12}^{(1)}-B_{22}^{(1)}(B_{11}^{(0)}+\lambda_i^{(0)})-
B_{11}^{(1)}(B_{22}^{(0)}+\lambda_i^{(0)})\right]\nonumber\\
& & +(B_{11}^{(0)}+\lambda_i^{(0)})(B_{22}^{(0)}+\lambda_i^{(0)})\left[(A_{21}^{(0)}-B_{21}^{(0)})
(A_{12}^{(1)}-B_{12}^{(1)})+(B_{22}^{(1)}-A_{22}^{(1)})(A_{11}^{(0)}-B_{11}^{(0)}-\lambda_i^{(0)})\right.
\nonumber\\
& & \left.
+(A_{11}^{(1)}-B_{11}^{(1)})(B_{22}^{(0)}-A_{22}^{(0)}+\lambda_i^{(0)})\right],
\end{eqnarray}
where
\begin{equation}
\label{a16}
A_{22}^{(1)}=\frac{1}{d+2}\mu_{12}^2(1+\alpha_{12})^2.
\end{equation}
\end{widetext}

\vicente{With respect to the shear stress $P_{xy}^*$, in the disordered phase ($\lambda_2^{(0)}>\lambda_1^{(0)}$ and so, $p_1^*=0$) one has $P_{1,xy}^*=0$ while $P_{2,xy}^*$ is given by (see Eq.\ \eqref{a11})
\begin{equation}
\label{a17}
P_{2,xy}^*=-\frac{A_{22}^{(0)}}{(B_{22}^{(0)}+\lambda_2^{(0)})^2}a^*.
\end{equation}}
\vicente{
In the ordered phase ($\lambda_1^{(0)}>\lambda_2^{(0)}$ and so, $p_1^*=\text{finite}$), Eqs.\ \eqref{a10} and \eqref{a11} yield, respectively,
\begin{equation}
\label{a19}
P_{1,xy}^*=\frac{d}{2a^*}\left(A_{11}^{(0)}-B_{11}^{(0)}-\lambda_1^{(0)}\right)p_1^*,
\end{equation}
%\begin{widetext}
\begin{eqnarray}
\label{a18}
P_{2,xy}^*&=&\frac{d}{2a^*}\left[A_{21}^{(0)}-B_{21}^{(0)}-\left(B_{22}^{(0)}+\lambda_1^{(0)}-
A_{22}^{(0)}\right.\right.\nonumber\\
& & \left.\left.+A_{21}^{(0)}-B_{21}^{(0)}\right)(1-p_1^*)\right],
\end{eqnarray}
where $p_1^*$ is given by Eq.\ \eqref{31}. Consequently, according to Eq.\ \eqref{a12}, the non-linear shear viscosity $\eta^*$ in the disordered phase is
\begin{equation}
\label{a20}
\eta^*=\frac{A_{22}^{(0)}}{(B_{22}^{(0)}+\lambda_2^{(0)})^2},
\end{equation}
while in the ordered phase the result is
\begin{eqnarray}
\label{a21}
\eta^*&=&\frac{d}{2a^{*2}}\left\{B_{22}^{(0)}+\lambda_1^{(0)}-A_{22}^{(0)}
-\left[\frac{d+2}{d}A_{21}^{(0)}\right.\right.\nonumber\\
& & \left.\left.
-d\left(B_{22}^{(0)}-A_{22}^{(0)}-B_{11}^{(0)}+A_{11}^{(0)}\right)\right]p_1^*\right\}.\nonumber\\
\end{eqnarray}}
%\end{widetext}


\begin{thebibliography} {99}
\bibitem{rqueref}
\emmanuel{Newtonian behaviour would correspond to a shear-rate independent temperature ratio,
to a vanishing normal stress difference ($P_{xx}=P_{yy}$), and to $P_{xy}$ proportional to
the shear rate (the proportionality factor being the viscosity).}
\bibitem{note}The value of the threshold $\mu_{\text{th}}$ for elastic collisions does not exactly coincide with the one obtained in Ref.\ \cite{MSG96} since, in the case of ordinary gases, the plain vanilla Maxwell model considered here is not completely equivalent to the conventional Maxwell model.
\bibitem{rque20}
As already discussed in the context of mixtures of ordinary gases \cite{MSG96},
the different qualitative behavior
of the system (ordered vs disordered) can be viewed in
analogy with the transition in a ferromagnetic
system \cite{S71}: The shear rate plays the role of the (inverse) temperature, the
energy ratio $E_1/E$ plays the role of the magnetization, and the mole fraction $x_1$
plays the role of the external magnetic field. One can then wonder what is the symmetry broken
for our system, to which the order parameter would be associated.
Given that near equilibrium $E_1/E \propto x_1$, then $x_1^{-1}E_1/E$
is invariant under the change $x_1\to \beta x_1$, where $\beta$ is an arbitrary constant.
This invariance is broken in the ordered phase since the mean kinetic energy
of the impurity is very much larger than the one corresponding to the excess particles.
\end{thebibliography}
\end{document}